\documentclass{jfm}
\usepackage{graphicx}
\usepackage{epstopdf, epsfig}
\usepackage{ar}
\usepackage[utf8]{inputenc}

\shorttitle{Energy transfer and resolvent modelling}
\shortauthor{S. Symon, S. J. Illingworth, and I. Marusic}

\title{Energy transfer in turbulent channel flows and implications for resolvent modelling}

\author{Sean Symon\aff{1}
	\corresp{\email{ssymon@unimelb.edu.au}},
	Simon J. Illingworth\aff{1}
	\and Ivan Marusic\aff{1}}

\affiliation{\aff{1}Department of Mechanical Engineering, University of Melbourne,
	Parkville, VIC 3010, Australia}

\begin{document}

\maketitle

\begin{abstract}
We analyse the inter-scale transfer of energy for two types of plane Poiseuille flow: the P4U exact coherent state of \citet{Park15} and turbulent flow in a minimal channel. For both flows, the dominant energy-producing modes are streamwise-constant streaks with a spanwise spacing of approximately 100 wall units. Since the viscous dissipation for these scales is not sufficient to balance production, the nonlinear terms redistribute the excess energy to other scales. Spanwise-constant scales (that is, Tollmien-Schlichting-like modes with zero spanwise wavenumber), in particular, account for a significant amount of net energy gain from the nonlinear terms. We compare the energy balance to predictions from resolvent analysis and we show that it does not model energy transfer well. Nevertheless, we find that the energy transferred from the streamwise-constant streaks can be predicted reasonably well by a Cess eddy viscosity profile. As such, eddy viscosity is an effective model for the nonlinear terms in resolvent analysis and explains good predictions for the most energetic streamwise-constant streaks. It also improves resolvent modes as a basis for structures whose streamwise lengths are greater than their spanwise widths by counteracting non-normality of the resolvent operator. Eddy viscosity does not respect the conservative nature of the nonlinear energy transfer which must sum to zero over all scales. It is less effective, consequently, for scales which receive energy from the nonlinear terms. 
\end{abstract}


\section{Introduction}

Energy transfer plays a key role in the organisation and evolution of turbulent flows. It is responsible for the multi-scale nature of turbulence through the Richardson-Kolmogorov turbulent energy cascade \citep{Kolmogorov41} and lends insight into the self-sustaining process \citep{Hamilton95}. Energy transfer for an individual scale is described by the spectral turbulent kinetic energy (TKE) equation, which contains a nonlinear term sometimes referred to as turbulent transport. As noted by \cite{Domaradzki94}, the nonlinearity poses considerable theoretical difficulties by permitting inter-scale energy exchange. It is not possible, for example, to study a scale in isolation without a closure model and, in the context of a large eddy simulation (LES), subgrid models need to account for the influence of small scales on the large scales of interest. An improved understanding of nonlinear interactions in turbulent flows, therefore, is essential to improve turbulence modelling and simulation. 

It is also known that linear mechanisms are important in energy transfer. These are described well by the linear operator obtained after linearising the Navier-Stokes equations around a suitable base flow \citep{Schmid01}. This operator is highly non-normal due to the mean shear found in wall-bounded flows \citep{Trefethen93}. As a result, infinitesimal disturbances may experience significant transient growth by extracting energy from the mean shear \citep{Butler92,Reddy93}. Linear mechanisms have also been identified in mean (time-averaged) flows by the resolvent analysis of \cite{McKeon10}. In this framework, the equations are linearised around the mean flow to obtain the resolvent operator that maps the nonlinear terms, treated as an intrinsic forcing, to the velocity in the frequency domain. Energy production by the linear system is excited by the nonlinear forcing which, when large enough, also excites dissipative modes to dissipate energy \citep{Sharma09}. The nonlinear forcing is itself composed of quadratic interactions between various outputs of the linear amplification process to complete a feedback loop \citep{McKeon13}. Unless the flow is dominated by a single Fourier mode \citep{Rosenberg19b}, it is not tractable to isolate the principal interactions that comprise the nonlinear forcing. 

An objective of this paper, therefore, is to investigate the extent to which energy transfer is correctly modelled by resolvent analysis. To address this question, we first examine how energy is produced, dissipated, and transferred among various scales in turbulent channel flow at low Reynolds numbers. Similar to other studies \citep{Mizuno16,Cho18,Lee19}, we calculate these terms in spectral space and integrate them over the wall-normal direction. The true energy transfer is compared to predictions from the optimal resolvent mode, which is often representative of the true velocity field observed in DNS or experiments \citep{McKeon17}. The agreement can be improved by adding the \cite{Cess58} eddy viscosity profile to the resolvent operator as done in many studies \citep{Hwang10,Morra19,Symon20}. It has been likened to a crude model for the energy cascade by \cite{Hwang16}, suggesting that it assumes the role of turbulent transport in resolvent analysis. To provide insight into the matter, we  quantify the contribution of eddy viscosity to the energy balance for each scale and compare it to the true nonlinear transfer. Finally, we aim to shed insight into the role of non-normality in the energy balance and the attenuation of non-normality by eddy viscosity to improve predictions from resolvent analysis.  

The flows selected for this study are the P4U exact coherent state (ECS) of \cite{Park15} and turbulent flow in a minimal channel at low Reynolds number. The former is a nonlinear travelling wave whose mean properties resemble those of near-wall turbulence. It is a particularly appealing choice to study energy transfer since it is low-dimensional \citep{Sharma16, Rosenberg19} and travels at a fixed convection velocity. As such, all computations can be performed on a standard personal computer and no integration in time is necessary to obtain each term in the energy budget. To verify that the transfer mechanisms are similar in a time-evolving flow, we compare the results for the P4U ECS to those of more standard turbulence in a ``minimal flow unit'' \citep{Jimenez91}. 

The paper is organised as follows. In \S\ref{sec:methods}, the relevant equations for resolvent analysis, energy transfer and the eddy viscosity model are derived. The simulation parameters for the P4U ECS and minimal channel flows are described in \S\ref{sec:description}. The energy balances computed from DNS and resolvent analysis are compared for the ECS in \S\ref{sec:ECS} and the minimal channel in \S\ref{sec:channel}. In \S\ref{sec:discussion}, we examine the influence of non-normality on the ability of the first resolvent mode to describe energy transfer processes. We also analyse the role of eddy viscosity on the efficiency of resolvent modes as a basis for the velocity fluctuations. This leads to a discussion on the typical scales for which an eddy viscosity leads to an improvement before we conclude in \S\ref{sec:conclusions}. 

\section{Methods} \label{sec:methods}

In \S\ref{sec:equations}, we describe the governing equations for plane Poiseuille flow and their non-dimensionalisation. A brief overview of resolvent analysis is provided in \S\ref{sec:resolvent analysis}. The energy balance for each scale is then derived from the fluctuation equations in \S\ref{sec:energy} and we show that this balance is maintained for each resolvent mode. Finally, we describe the eddy viscosity model in \S\ref{sec:eddy viscosity}.

\subsection{Plane Poiseuille  flow equations} \label{sec:equations}

The non-dimensional Navier-Stokes equations for statistically steady, turbulent plane Poiseuille flow are
\begin{subequations} \label{eq:NSE}
	\begin{equation} \label{eq:Momentum}
	\frac{\partial \boldsymbol{u}}{\partial t} + \boldsymbol{u} \cdot \boldsymbol{\nabla} \boldsymbol{u}  = -\boldsymbol{\nabla} p + \frac{1}{Re_{\tau}}\boldsymbol{\nabla}^2\boldsymbol{u},	
	\end{equation} 
	\begin{equation}
	\boldsymbol{\nabla} \cdot \boldsymbol{u} = 0,
	\end{equation}
\end{subequations}
where $\boldsymbol{u} = [u,v,w]^T$ is the velocity in the $x$ (streamwise), $y$ (spanwise) and $z$ (wall-normal) directions and $p$ is the pressure. The friction Reynolds number $Re_{\tau} = u_\tau h /\nu$ is defined in terms of the friction velocity $u_{\tau}$, channel half height $h$, and kinematic viscosity $\nu$. No-slip boundary conditions are applied at the walls and periodic boundary conditions are imposed in the streamwise and spanwise directions. The density of the fluid is $\rho$ and the velocities are non-dimensionalized by $u_\tau$, the spatial variables by $h$ and the pressure by $\rho u_{\tau}^2$. A `$+$' superscript denotes spatial variables that have been normalized by the viscous length scale $\nu/u_{\tau}$. 

\subsection{Resolvent analysis} \label{sec:resolvent analysis}

We begin by Reynolds-decomposing (\ref{eq:Momentum}), which leads to the following equations for the fluctuations:
\begin{equation} \label{eq:input-output}
\frac{\partial \boldsymbol{u}'}{\partial t} + \boldsymbol{U} \cdot \boldsymbol{\nabla} \boldsymbol{u}' + \boldsymbol{u}' \cdot \boldsymbol{\nabla} \boldsymbol{U} + \boldsymbol{\nabla} p' - \frac{1}{Re_{\tau}}\boldsymbol{\nabla}^2 \boldsymbol{u}' = -\boldsymbol{u}'\cdot \boldsymbol{\nabla} \boldsymbol{u}' + \overline{\boldsymbol{u}' \cdot \boldsymbol{\nabla} \boldsymbol{u}'} = \boldsymbol{f}',
\end{equation}
where $(\overline{\cdot})$ and $(\cdot)'$ denote a time-average and fluctuation, respectively, and $\boldsymbol{U} = [U(y),0,0]^T$ is the mean velocity. Equation (\ref{eq:input-output}) is written such that all linear terms appear on the left-hand side while all nonlinear terms appear on the right-hand side. Equation (\ref{eq:input-output}) is then Fourier-transformed in time and in the homogeneous directions $x$ and $y$
\begin{equation} \label{eq:Fourier transform}
\hat{\boldsymbol{u}}(k_x,k_y,z,\omega) = \int_{\infty}^{\infty}\int_{-\infty}^{\infty} \int_{-\infty}^{\infty} \boldsymbol{u}'(x,y,z,t) e^{-i(k_xx+k_yy-\omega t)}dxdydt,
\end{equation}
where $(\hat{\cdot})$ denotes the Fourier-transformed coefficient, $k_x$ is the streamwise wavenumber, $k_y$ is the spanwise wavenumber and $\omega$ the temporal frequency. The equivalent wavelengths in the streamwise and spanwise directions are $\lambda_x = 2\pi/k_x$ and $\lambda_y = 2\pi/k_y$. The equations are arranged into state-space form \citep{Jovanovic05} after substituting (\ref{eq:Fourier transform}) into (\ref{eq:input-output})
\begin{subequations} \label{eq:OSSQ}
	\begin{equation} 
	i\omega \hat{\boldsymbol{q}}(\boldsymbol{k}) = \boldsymbol{A}(k_x,k_y)\hat{\boldsymbol{q}}(\boldsymbol{k})  + \boldsymbol{B}(k_x,k_y)\hat{\boldsymbol{f}}(\boldsymbol{k}) ,
	\end{equation} 
	\begin{equation}
	\hat{\boldsymbol{u}}(\boldsymbol{k})  = \boldsymbol{C}(k_x,k_y)\hat{\boldsymbol{q}}(\boldsymbol{k}) , 
	\end{equation}
\end{subequations}
where $\boldsymbol{k} = (k_x,k_y,\omega)$ is the wavenumber triplet and $\hat{\boldsymbol{q}}$ consists of the wall-normal velocity and vorticity $\hat{\eta} = ik_y\hat{u} - ik_x\hat{v}$. The operators $\boldsymbol{A}$, $\boldsymbol{B}$ and $\boldsymbol{C}$ represent the linear Navier-Stokes operator, the input matrix and the output matrix, respectively, and are defined in appendix \ref{sec:operators}. These operators are independent of $\omega$ but are functions of the wavenumber pair $(k_x,k_y)$ under consideration. In the interest of readability, this dependence is omitted for the rest of the paper. 

Once (\ref{eq:OSSQ}) is recast into input-output form, i.e, 
\begin{equation}
\hat{\boldsymbol{u}}(\boldsymbol{k}) = \boldsymbol{C}(i\omega \boldsymbol{I}- \boldsymbol{A})^{-1} \hat{\boldsymbol{f}}(\boldsymbol{k}) = \mathcal{H}(\boldsymbol{k}) \hat{\boldsymbol{f}}(\boldsymbol{k}),
\end{equation}
a linear operator called the resolvent $\mathcal{H}(\boldsymbol{k})$ relates the input forcing $\hat{\boldsymbol{f}}(\boldsymbol{k})$ to the output velocity $\hat{\boldsymbol{u}}(\boldsymbol{k})$. Even if the nonlinear forcing is unknown, the resolvent identifies structures due to linear mechanisms. These structures can be obtained from a singular value decomposition of the resolvent operator:
\begin{equation} \label{eq:SVD}
\mathcal{H}(\boldsymbol{k}) = \hat{\boldsymbol{\Psi}} (\boldsymbol{k}) \boldsymbol{\Sigma}(\boldsymbol{k}) \hat{\boldsymbol{\Phi}}^*(\boldsymbol{k}),
\end{equation} 
where $\hat{\boldsymbol{\Psi}}(\boldsymbol{k}) = [\hat{\boldsymbol{\psi}}_1(\boldsymbol{k}), \hat{\boldsymbol{\psi}}_2(\boldsymbol{k}), \cdots, \hat{\boldsymbol{\psi}}_p(\boldsymbol{k})]$ and  $\hat{\boldsymbol{\Phi}}(\boldsymbol{k}) = [\hat{\boldsymbol{\phi}}_1(\boldsymbol{k}),\hat{\boldsymbol{\phi}}_2(\boldsymbol{k}), \cdots, \hat{\boldsymbol{\phi}}_p(\boldsymbol{k})]$ are orthogonal basis functions for the velocity and nonlinear forcing, respectively. The diagonal matrix $\boldsymbol{\Sigma}(\boldsymbol{k})$ ranks the $p$th structure by its gain $\sigma_p(\boldsymbol{k})$ using an inner product that is proportional to its kinetic energy. Consequently, the structure $\hat{\boldsymbol{\psi}}_1(\boldsymbol{k})$, referred to as the optimal or first resolvent mode, is the most amplified response by the linear dynamics contained in the operator. The true velocity field is the weighted sum of resolvent modes, i.e.,
\begin{equation} \label{eq:weighted sum}
\hat{\boldsymbol{u}}(\boldsymbol{k}) = \sum_{p = 1}^N \hat{\boldsymbol{\psi}}_p(\boldsymbol{k}) \sigma_p(\boldsymbol{k})  \chi_p(\boldsymbol{k}),
\end{equation}
where $\chi_p(\boldsymbol{k})$ is the projection of $\hat{\boldsymbol{\phi}}_p(\boldsymbol{k})$ onto $\hat{\boldsymbol{f}}(\boldsymbol{k})$. 

\subsection{Energy balance} \label{sec:energy}

We now derive the energy balance that must be satisfied by the velocity field and individual resolvent modes. Equation (\ref{eq:input-output}) is rewritten in index notation
\begin{equation} \label{eq:fluctuating NSE}
\frac{\partial u'_i}{\partial t}  + U_j\frac{\partial u'_i}{\partial x_j} + u_j'\frac{\partial U_i}{\partial x_j}  + \frac{\partial p'}{\partial x_i} - \frac{1}{Re}\frac{\partial^2u_i'}{\partial x_j \partial x_j} = - u'_j\frac{\partial u_i'}{\partial x_j} + \overline{u'_j\frac{\partial u_i'}{\partial x_j}} .
\end{equation}
Similar to (\ref{eq:input-output}), all nonlinear terms appear on the right-hand side although they are not treated as an unknown forcing. The indices $i,j =1,2,3$ and $U_i = (U(z),0,0)$ is the mean velocity, which is a function of the wall-normal direction only. It can be noted, therefore, that $U_1 = U$, $U_j = 0$ if $j=2,3$ and $\partial U_i/\partial x_j \neq 0$ for $i = 1$ and $j = 3$ only. The kinetic energy of the full system is characterised by the inner product between (\ref{eq:fluctuating NSE}) and $u_i'$ integrated over the volume $V$:
\begin{eqnarray} \label{eq:Reynolds-Orr}
\underbrace{ \frac{1}{2} \int_V \frac{\partial u_i^{'2}}{\partial t} dV }_{\dot{E}(t)} = \underbrace{- \int_V u'_i u_j' \frac{\partial U_i}{\partial x_j} dV }_{P(t)} \underbrace{ - \frac{1}{Re} \int_V \frac{\partial u_i'}{\partial x_j} \frac{\partial u_i'}{\partial x_j} dV}_{D(t)},\\
\int_V u_i'f_i'dV = \int_V u_i'u_j'\frac{\partial u_i'}{\partial x_j}dV = 0. \label{eq:conservative}
\end{eqnarray}
Equation (\ref{eq:Reynolds-Orr}) is the Reynolds-Orr equation \citep{Schmid01} where the evolution of kinetic energy in the system is a balance between production and dissipation, which must be negative. Due to the conservative nature of the nonlinear terms, their contribution to the Reynolds-Orr equation sums to zero when integrated over the volume as expressed in (\ref{eq:conservative}). For a statistically stationary flow, a time average of (\ref{eq:Reynolds-Orr}) implies that production balances dissipation since $\overline{dE/dt} = 0$. 

The kinetic energy for a specific spatial scale is obtained after multiplying (\ref{eq:fluctuating NSE}) by $u_i^{'*}$ and Fourier-transforming in $x$ and $y$. The result is integrated over the wall-normal direction and time-averaged to arrive at the spectral turbulent kinetic energy (TKE) equation:
\begin{eqnarray} \label{eq:budget}
\nonumber \overline{\frac{\partial  \hat{E}(k_x,k_y)}{\partial t}} =  & \underbrace{- \int_{-h}^h\frac{dU}{dz}\overline{\hat{u}^*(k_x,k_y)\hat{v}(k_x,k_y)}dz}_{\hat{P}(k_x,k_y)}  \underbrace{-\frac{1}{Re}\int_{-h}^h \overline{ \frac{\partial \hat{u}_i(k_x,k_y)}{\partial x_j} \frac{\partial \hat{u}^*_i(k_x,k_y)}{\partial x_j}} dz}_{\hat{D}(k_x,k_y)} \\ &
 \underbrace{ - \int_{-h}^h \overline{\hat{u}_i^*(k_x,k_y)\frac{\partial}{\partial x_j} \widehat{u_iu_j}(k_x,k_y) }  dz}_{\hat{N}(k_x,k_y)} = 0.
\end{eqnarray}
The pressure terms vanish in (\ref{eq:budget}) after integrating over the channel height \citep{Aubry88}. Following \cite{Muralidhar19}, we consider the real part of (\ref{eq:budget}), which consists of three terms: production, viscous dissipation and nonlinear transfer. In general, production $\hat{P}$ is positive for a given scale as perturbations extract energy from the mean flow. Viscous dissipation $\hat{D}$, on the other hand, is guaranteed to be real and negative according to (\ref{eq:budget}) as it is the mechanism through which kinetic energy is removed from the system and converted into heat. Nonlinear transfer $\hat{N}$ may be positive or negative depending on the scale selected. If $\hat{P} > \hat{D}$, for example, then $\hat{N} < 0$ in order to achieve a balance. In a similar fashion, if $\hat{P} < \hat{D}$, then $\hat{N} > 0$. The integral of $\hat{N}$ over all $k_x$ and $k_y$, nevertheless, is zero as stated in (\ref{eq:conservative}).

To obtain the energy balance for resolvent modes, which are defined for a wavenumber triplet $\boldsymbol{k}$, (\ref{eq:fluctuating NSE}) is multiplied by $u_i^{'*}$ and Fourier-transformed in $x$, $y$ and $t$. The result is integrated over the wall-normal direction 
\begin{equation} \label{eq:budget k}
 \int_{-h}^h\frac{dU}{dz} \hat{u}^*(\boldsymbol{k})\hat{v} (\boldsymbol{k})dz 
-\frac{1}{Re} \int_{-h}^h \frac{\partial \hat{u}_i(\boldsymbol{k})}{\partial x_j} \frac{\partial \hat{u}^*_i(\boldsymbol{k})}{\partial x_j} dz 
- \int_{-h}^h \hat{u}_i^*(\boldsymbol{k})\hat{f}_i(\boldsymbol{k}) dz = 0.
\end{equation}
In this form, the nonlinear forcing $\hat{\boldsymbol{f}}(\boldsymbol{k})$ appears explicitly in the energy balance. We can now express the velocity field in terms of resolvent modes. In the special case where $\hat{\boldsymbol{f}}(\boldsymbol{k})$ is white noise, the velocity field can be written as
\begin{equation} \label{eq:white noise resolvent}
\hat{\boldsymbol{u}}(\boldsymbol{k}) = \sum_{p = 1}^N \hat{\boldsymbol{\psi}}_p(\boldsymbol{k}) \sigma_p(\boldsymbol{k}).
\end{equation}
Substituting (\ref{eq:white noise resolvent}) into (\ref{eq:budget k}) yields 
\begin{eqnarray} \label{eq:res mode balance}
 \nonumber \sum_p  \sigma_p(\boldsymbol{k}) \left( \int_{-h}^h \frac{dU}{dz} \hat{\boldsymbol{\psi}}_p^{*,i=1}(\boldsymbol{k}) \hat{\boldsymbol{\psi}}_p^{j=2} (\boldsymbol{k}) dz + \frac{1}{Re} \int_{-h}^h \frac{\partial \hat{\boldsymbol{\psi}}_p^{*,i}(\boldsymbol{k})} {\partial \boldsymbol{x}_j} \frac{\partial \hat{\boldsymbol{\psi}}_p^i (\boldsymbol{k})}{\partial \boldsymbol{x}_j} dz \right) + \\ 
\sum_p \int_{-h}^h \boldsymbol{\psi}_p^*(\boldsymbol{k})\hat{\boldsymbol{\phi}}_p(\boldsymbol{k})dz = 0.
\end{eqnarray}
Each term in the sum can be decoupled since the basis functions $\hat{\boldsymbol{\psi}}_p$ are orthogonal. This means that production, dissipation and nonlinear transfer must be balanced across each resolvent mode. If $\sigma_1(\boldsymbol{k}) \gg \sigma_2(\boldsymbol{k})$ then it would imply that the sum over all $p$ is dominated by the first resolvent mode, or
\begin{eqnarray} \label{eq:budget rank 1}
\nonumber  \sigma_1(\boldsymbol{k}) \left( \int_{-h}^h \frac{dU}{dz} \hat{\boldsymbol{\psi}}_1^{*,i=1}(\boldsymbol{k}) \hat{\boldsymbol{\psi}}_1^{j=2} (\boldsymbol{k}) dz + \frac{1}{Re} \int_{-h}^h \frac{\partial \hat{\boldsymbol{\psi}}_1^{*,i}(\boldsymbol{k})} {\partial \boldsymbol{x}_j} \frac{\partial \hat{\boldsymbol{\psi}}_1^i (\boldsymbol{k})}{\partial \boldsymbol{x}_j} dz \right) + \\ 
 \int_{-h}^h \boldsymbol{\psi}_1^*(\boldsymbol{k})\hat{\boldsymbol{\phi}}_1(\boldsymbol{k})dz = 0.
\end{eqnarray}
The bulk of production, dissipation and nonlinear transfer for a particular scale $\boldsymbol{k}$, therefore, could also be accounted for by the first resolvent mode.  

\subsection{Eddy viscosity model} \label{sec:eddy viscosity}

If the nonlinear forcing is not white noise, then (\ref{eq:res mode balance}) is not applicable since it does not take into account the complex amplitude of each mode. One method to model the nonlinear forcing is to add an eddy viscosity to the linearised equations after performing a triple decomposition of the velocity field $\tilde{\boldsymbol{u}}$ into a mean component $\boldsymbol{U}$, coherent motions $\boldsymbol{u}$ and incoherent turbulent fluctuations $\boldsymbol{u}'$ \citep{Reynolds72}. The equations governing the coherent velocity and pressure are 
\begin{equation} \label{eq:eddy LNSE}
\frac{\partial \boldsymbol{u}}{\partial t} + \boldsymbol{U} \cdot \boldsymbol{\nabla} \boldsymbol{u} + \boldsymbol{u} \cdot \boldsymbol{\nabla} \boldsymbol{U} + \boldsymbol{\nabla} p + \boldsymbol{\nabla} \cdot \left[ \nu_T ( \boldsymbol{\nabla} \boldsymbol{\boldsymbol{u}} + \boldsymbol{\nabla} \boldsymbol{\boldsymbol{u}}^T) \right] = \boldsymbol{d},
\end{equation}
where $\nu_T(z)$ is the total effective viscosity and $\boldsymbol{d} = -\boldsymbol{u} \cdot \boldsymbol{\nabla} \boldsymbol{u} + \overline{\boldsymbol{u} \cdot \boldsymbol{\nabla} \boldsymbol{u}}$ is the forcing term. It should be noted that $\boldsymbol{d}$ is different from $\boldsymbol{f}'$ in (\ref{eq:input-output}). Following \cite{Reynolds67} and \cite{Hwang10}, we use the \cite{Cess58} eddy viscosity profile of the form 
\begin{equation}
\nu_T(z) = \frac{\nu}{2}\left(1 + \left[ \frac{\kappa}{3}(1-z^2)(1+2z^2)\left(1-e^{|z-1| \frac{Re_{\tau}}{A}}\right) \right]^2 \right)^{1/2} + \frac{\nu}{2},
\end{equation}
where $\kappa = 0.426$ and $A = 25.4$ are chosen based on a least-squares fit to experimentally obtained mean velocity profiles at $Re_{\tau} = 2000$ \citep{delAlamo06}. 

Fourier-transforming (\ref{eq:eddy LNSE}) in time and the homogeneous directions and rearranging it into the following input-output form yields
\begin{equation}
\hat{\boldsymbol{u}}(\boldsymbol{k}) = \mathcal{H}^e(\boldsymbol{k})\hat{\boldsymbol{d}}(\boldsymbol{k}),
\end{equation}
where $\mathcal{H}^e(\boldsymbol{k})$ is a new linear operator that relates the forcing $\hat{\boldsymbol{d}} (\boldsymbol{k})$ to the velocity field $\hat{\boldsymbol{u}}(\boldsymbol{k})$. Similar to the resolvent operator, we can analyse structures that are preferentially amplified by performing a singular value decomposition 
\begin{equation}
\mathcal{H}^e(\boldsymbol{k}) = \hat{\boldsymbol{\Psi}}^e (\boldsymbol{k}) \boldsymbol{\Sigma}^e (\boldsymbol{k}) \hat{\boldsymbol{\Phi}}^{*,e}(\boldsymbol{k}),
\end{equation}
although the individual modes $\hat{\boldsymbol{\psi}}_p^e(\boldsymbol{k})$ do not satisfy the energy balance in (\ref{eq:res mode balance}). Instead, the addition of eddy viscosity in (\ref{eq:eddy LNSE}) introduces two terms, the first of which is
\begin{equation} \label{eq:eddy dissipation}
\hat{V}(k_x,k_y) =  -\int_{-h}^h (\nu_T(z)-\nu) \overline{ \frac{\partial \hat{u}_i(k_x,k_y)}{\partial x_j} \frac{\partial \hat{u}^*_i(k_x,k_y)}{\partial x_j}} dz, 
\end{equation}
where the kinematic viscosity $\nu$ has been subtracted in order to remove the contribution of viscous dissipation $\hat{D}(k_x,k_y)$. The remainder $\hat{V}(k_x,k_y)$ represents, therefore, the additional dissipation provided by the wall-normal varying portion of $\nu_T$. Similar to $\hat{D}(k_x,k_y)$, this term is real and negative, signifying that it removes energy. The second term is related to the wall-normal gradient of $\nu_T$
\begin{equation}
\hat{G}(k_x,k_y) =  -\int_{-h}^h \frac{d \nu_T}{dz} \overline{ \left( \hat{u}^*_i(k_x,k_y) \frac{\partial \hat{u}_i(k_x,k_y)}{\partial y} + \hat{u}_i^*(k_x,k_y) \frac{\partial \hat{v}(k_x,k_y)}{\partial x_i} \right) } dz.
\end{equation}
Unlike $\hat{V}(k_x,k_y)$, the sign of $\hat{G}(k_x,k_y)$ cannot be determined \textit{a priori}. 

The combined effect of $\hat{V}(k_x,k_y)$ and $\hat{G}(k_x,k_y)$ is referred to as eddy dissipation $\widehat{Edd}(k_x,k_y)$, i.e.,
\begin{equation}
\widehat{Edd}(k_x,k_y) = \hat{V}(k_x,k_y) + \hat{G}(k_x,k_y).
\end{equation}
Eddy dissipation is computed in \S\S\ref{sec:ECS} and \ref{sec:channel} using the true velocity field to determine its accuracy in modelling the effect of nonlinear transfer in (\ref{eq:budget}). If $\widehat{Edd}(k_x,k_y) \approx \hat{N}(k_x,k_y)$, then it is expected that eddy viscosity will lead to an improvement in the structures predicted by resolvent analysis.

\section{Flow descriptions} \label{sec:description}

In this section, we describe the two flows that are analysed from an energy transfer perspective. These are the P4U ECS computed by \cite{Park15} and turbulent channel flow in the minimal unit \citep{Jimenez91} which are discussed in \S\S\ref{sec:ECS description} and \ref{sec:min chan description}, respectively.

\subsection{P4U ECS} \label{sec:ECS description}

\begin{table}
	\begin{center}
		\def~{\hphantom{0}}
		\begin{tabular}{lccccccc}
			& $Re_{\tau}$  & $ c^+$ & $L_x$ & $L_y$ & $N_x$ & $N_y$ & $N_z$     \\[3pt]
			P4U  & 85 & 14.2 & $\pi$ & $\pi/2$ & 24 & 24 & 81 \\
			Minimal Channel & 180 & $ \in [0,19.4]$ &$\pi$ & $\pi/4$ & 96 & 48 & 128  \\
		\end{tabular}
		\caption{Relevant parameters for the flows under consideration.}
		\label{tab:parameters}
	\end{center}
\end{table}

\begin{figure}
	\centering
	\includegraphics[trim = 0cm 0cm 1cm 0cm, clip,scale=0.35]{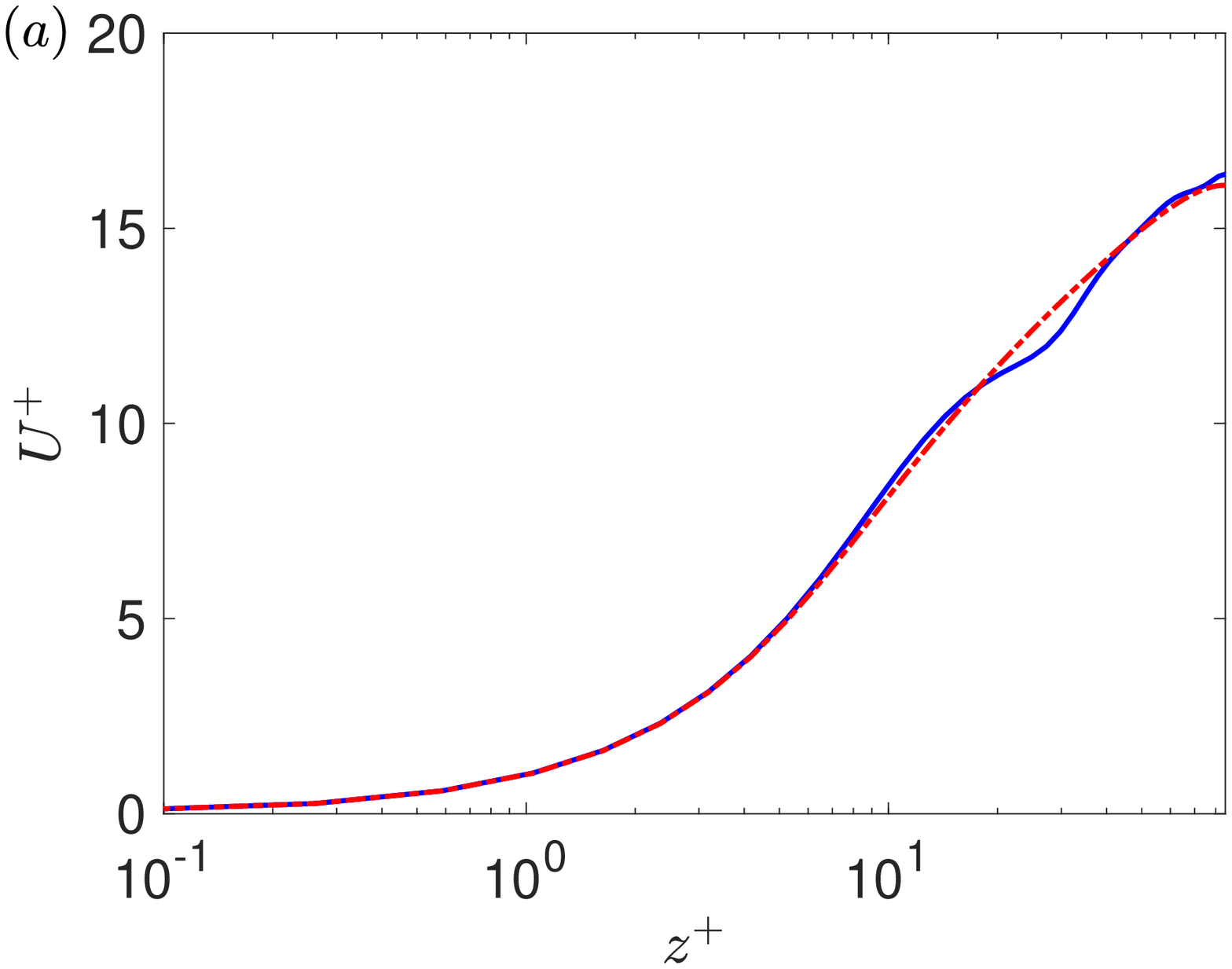} 
	\includegraphics[trim = 0cm 0cm 1cm 0cm, clip,scale=0.35]{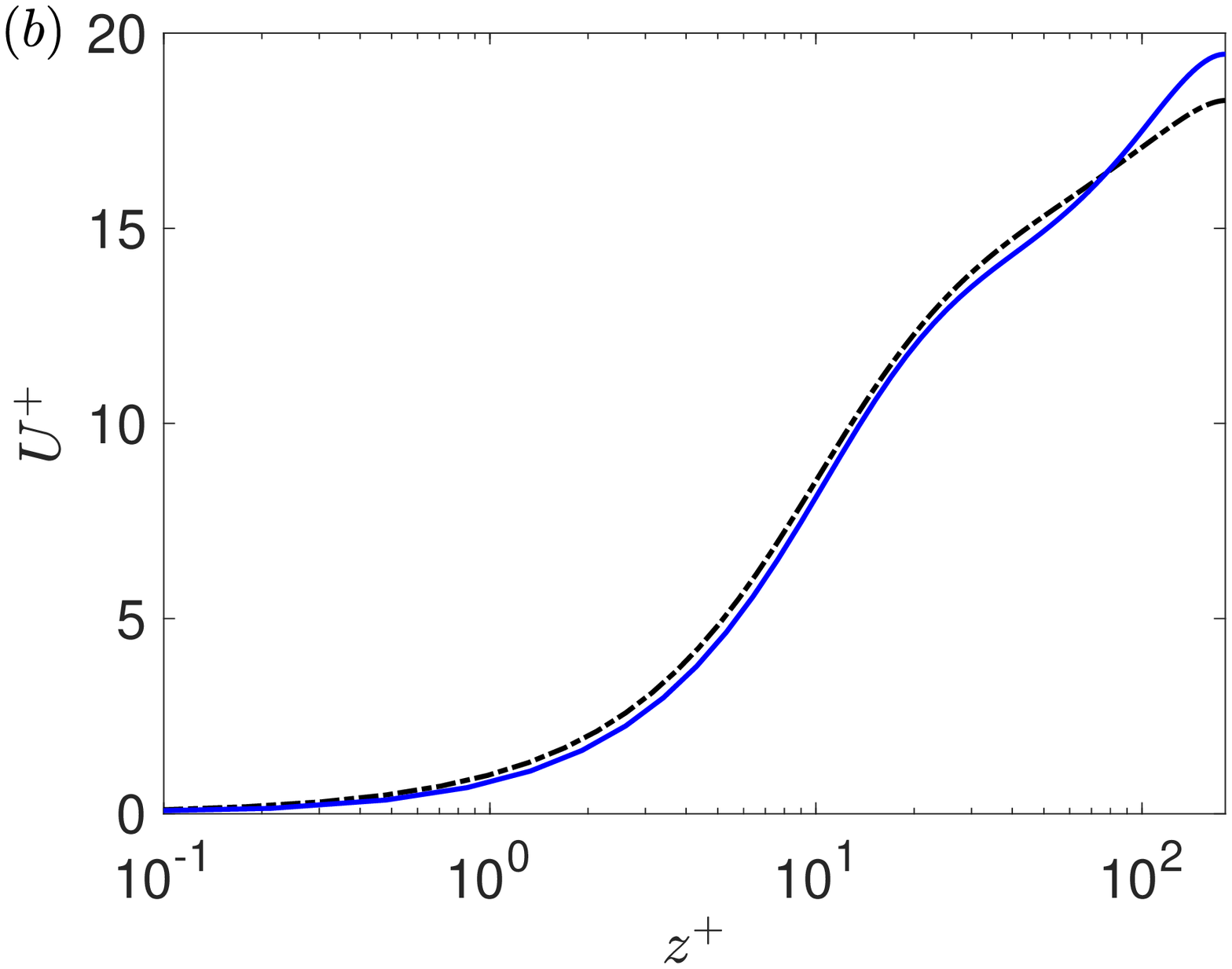}
	
	\caption{(a) Mean velocity profiles for P4U (blue) and Cess model (red) at $Re_{\tau} = 85$. (b) Mean velocity profiles for the minimal channel (blue) and the DNS of \cite{Lee15} (black). }\label{fig:means}
\end{figure}

The P4U ECS, henceforth referred to as P4U, is a nonlinear travelling wave with a friction Reynolds number of $Re_{\tau} = 85$ and fixed wave speed of $c^+ = 14.2$. As seen in table~\ref{tab:parameters}, P4U is solved in a computational domain with 24 equally spaced grid points in the streamwise and spanwise directions, which have lengths of $\pi$ and $\pi/2$, respectively. There are 81 points in the wall-normal direction on a Chebyshev grid. The spatial structure of P4U is in the form of low-speed streaks, which are wavy in the streamwise direction, straddled by counter-rotating vortices. As mentioned by \cite{Park15}, its structure is qualitatively similar to near-wall turbulence and its mean velocity profile closely resembles a standard turbulent mean. This is seen more clearly in figure \ref{fig:means}(a) where the mean profile for P4U in blue is compared to the Cess model in red at $Re_{\tau} = 85$. Despite good overall agreement, the P4U ECS profile has a more wavy nature since the structure has a single convection velocity.

\begin{figure}
	\centering
	\includegraphics[trim = 1.8cm 0cm 3.2cm 0cm, clip,scale=0.3]{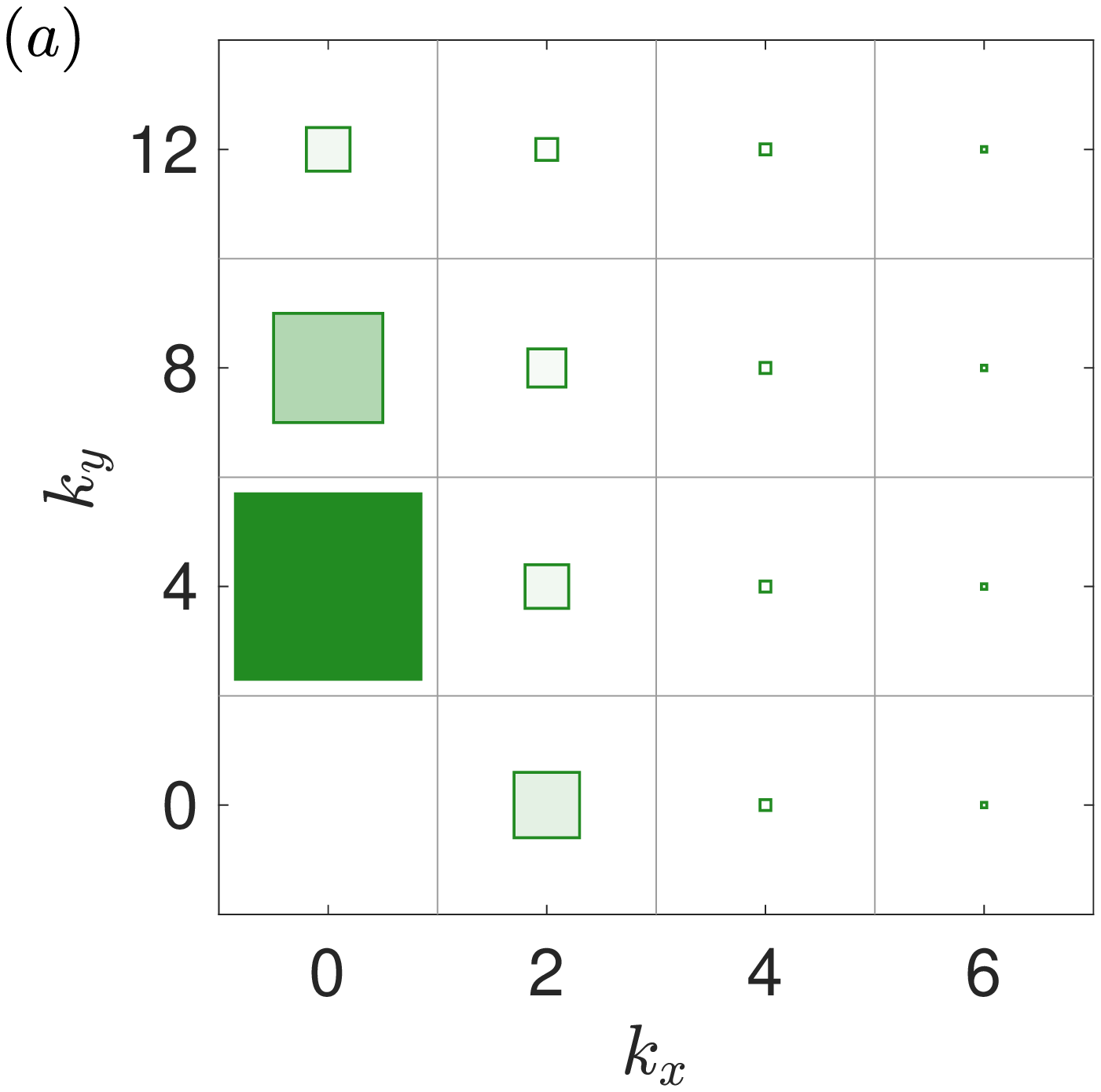}
	\includegraphics[trim = 1.8cm 0cm 3.2cm 0cm, clip,scale=0.3]{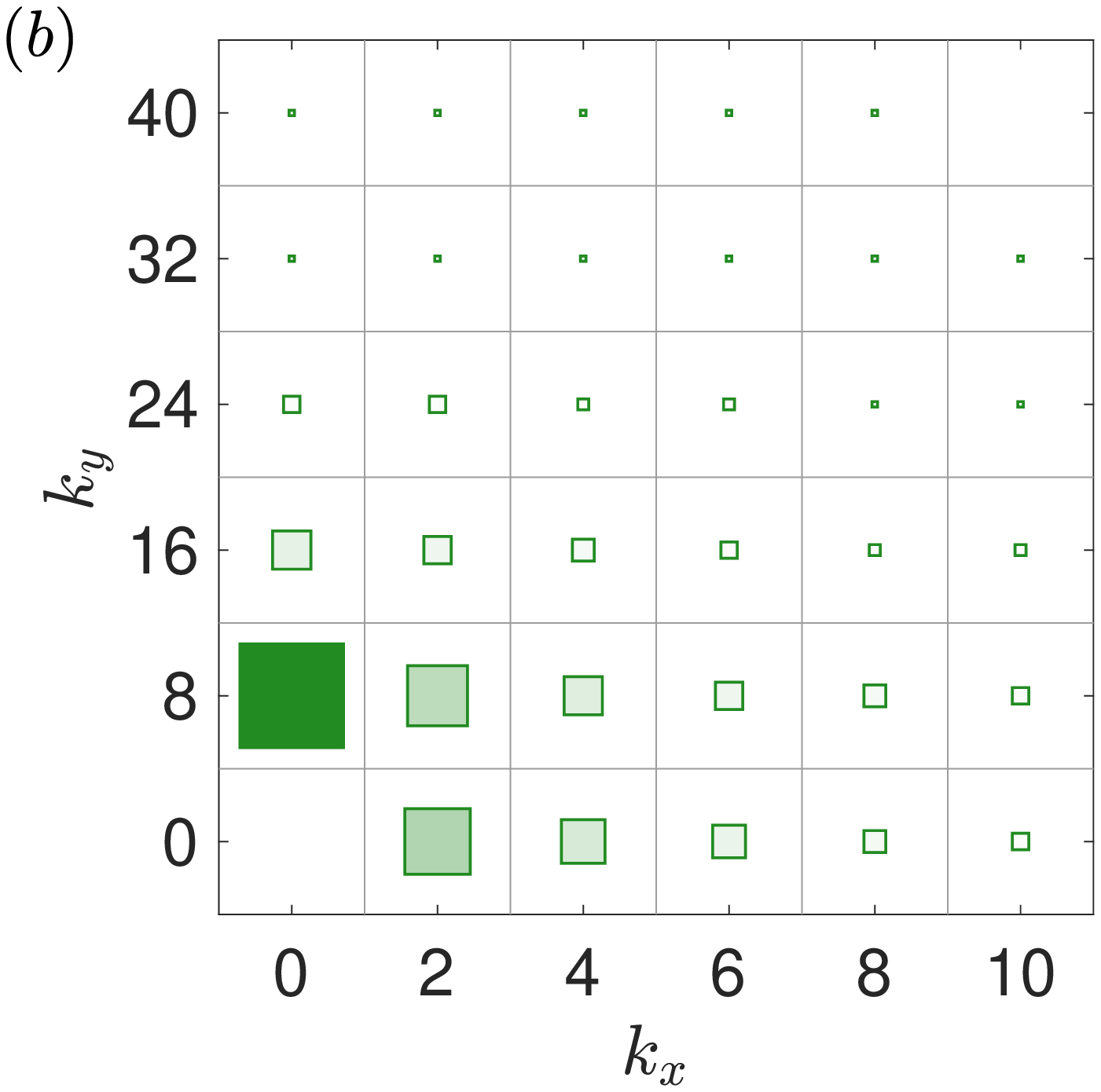}
	
	\caption{The kinetic energy $\hat{E}(k_x,k_y)$ of the most energetic wavenumber pairs in (a) P4U and the (b) minimal channel. The area and colour intensity of the square marker at the centre of each tile are directly proportional to the kinetic energy.}\label{fig:energy}
\end{figure}

Even though the simulation size is small, there are still many wavenumber pairs which may participate in the transfer of energy. We begin by computing the kinetic energy of each wavenumber pair 
\begin{equation}
\hat{E}(k_x,k_y) = \frac{1}{2}\left(\overline{\hat{u}^2(k_x,k_y) + \hat{v}^2(k_x,k_y) + \hat{w}^2(k_x,k_y)}\right),
\end{equation}
and plot the most energetic pairs in figure \ref{fig:energy}(a). The area and colour intensity of the square marker at the centre of each tile are directly proportional to the kinetic energy. The most energetic scale is streamwise-constant with a spanwise width of approximately 100 wall units, which matches the near-wall streak spacing of \cite{Smith83}. Most of the kinetic energy, furthermore, is concentrated in structures with small streamwise wavenumbers.

\subsection{Minimal channel} \label{sec:min chan description}

The minimal channel flow is computed for $Re_{\tau} = 180$ using an unstructured finite difference solver (see \cite{Chung14} for details) on a domain with dimensions $\pi \times \pi/4 \times 2h$ in the streamwise, spanwise and wall-normal directions. There are 96 and 48 equally spaced points in the streamwise and spanwise directions, respectively, and 128 points in the wall-normal direction on a Chebyshev grid. The mean profile for the minimal channel in figure \ref{fig:means}(b) is more smooth than the P4U mean profile since there exists a distribution of wave speeds which range between $0 < c^+ < 19.4$ as seen in table \ref{tab:parameters}. The minimal channel mean profile is in good agreement with the mean profile of \cite{Lee15} for most areas of the flow other than the wake region, where the minimal channel mean profile overshoots the one from \cite{Lee15}. This phenomenon has been observed by \cite{Jimenez91} and stems from the fact that the minimal domain is too small to accommodate the largest structures which reside in the outer region. Despite this disagreement, there is no impact on near-wall turbulence in the buffer and viscous regions where the bulk of energy resides \citep{Jimenez91,Jimenez99}. 

Similar to P4U, the kinetic energy for the most energetic wavenumber pairs is plotted in figure \ref{fig:energy}(b). Although there are more energetic scales in the minimal channel since the friction Reynolds number is higher, the relative distribution of energy among the scales is quite similar to P4U. The most energetic scale is also streamwise-constant with a spanwise width of approximately 100 wall units. This supports the notion that P4U is a simple model for turbulent channel flow at very low Reynolds number. Therefore, to facilitate visualisation later in the paper, we choose to plot only those wavenumber pairs that appear in figure \ref{fig:energy} although the energy balance will be computed across all of them. 

\section{Results: P4U ECS} \label{sec:ECS}

In this section, we analyse energy transfer for P4U. We begin with a comparison of production, dissipation and nonlinear transfer across the most energetic scales in \S\ref{sec:P4 balance}. These results are compared to the resolvent predictions in \S\ref{sec:P4 resolvent}. Finally, the additional dissipation introduced by eddy viscosity is quantified for each scale and compared to nonlinear transfer in \S\ref{sec:P4 eddy}.

\subsection{Energy balance} \label{sec:P4 balance}

\begin{figure}
	\centering
	\includegraphics[trim = 1.8cm 0cm 3.2cm 0cm, clip,scale=0.3]{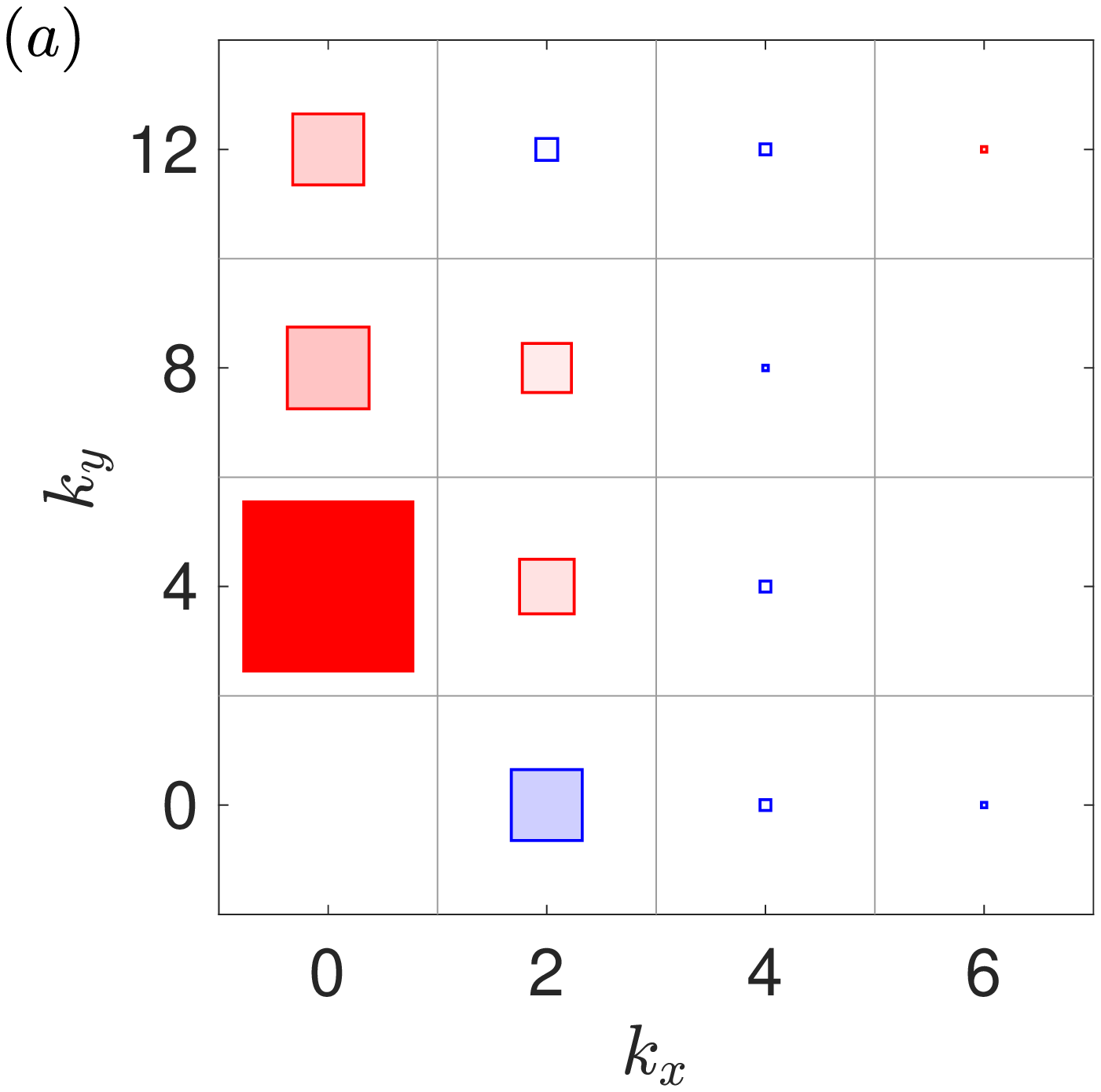}
	\includegraphics[trim = 1.8cm 0cm 3.2cm 0cm, clip,scale=0.3]{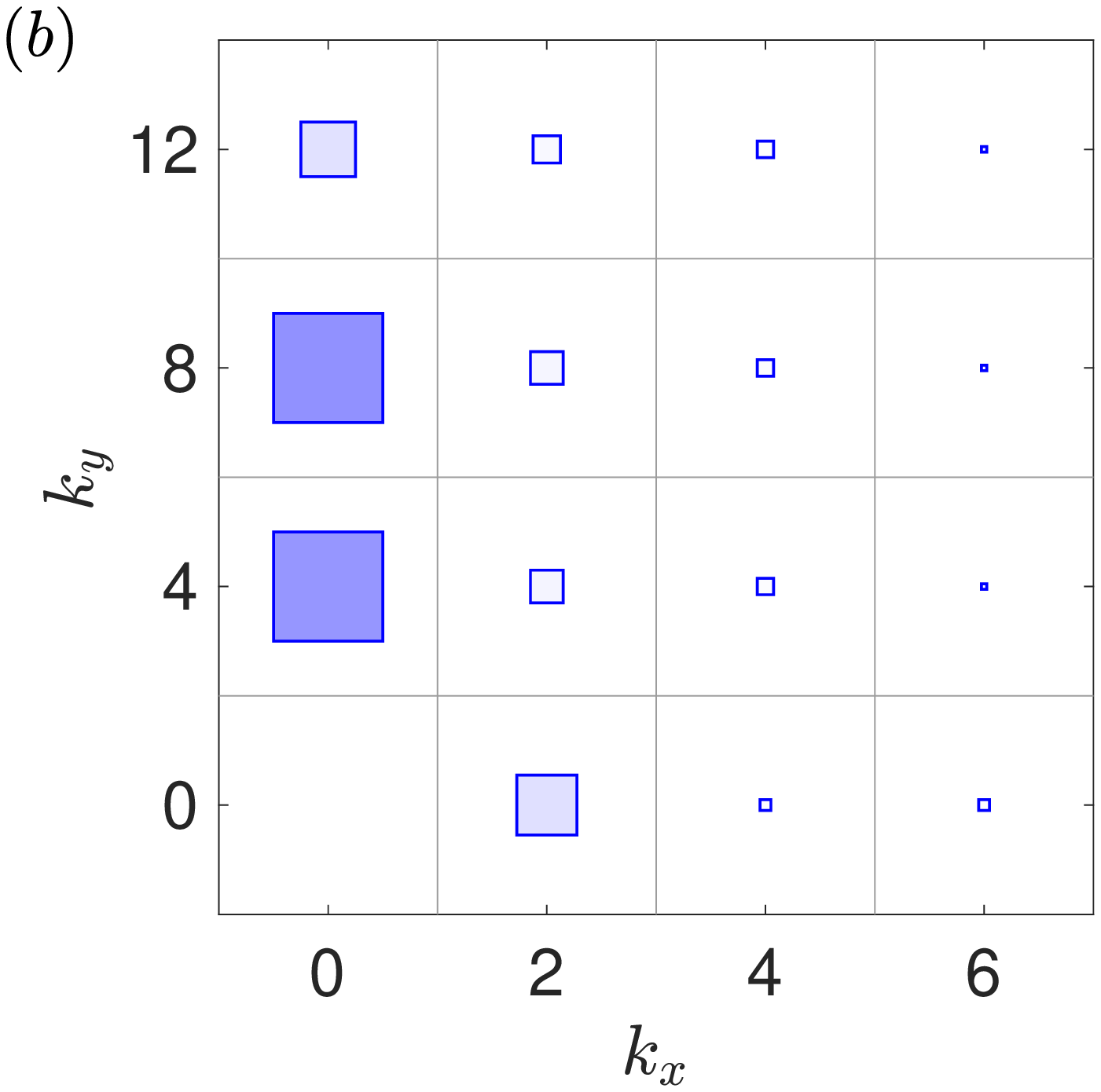}
	\includegraphics[trim = 1.8cm 0cm 3.2cm 0cm, clip,scale=0.3]{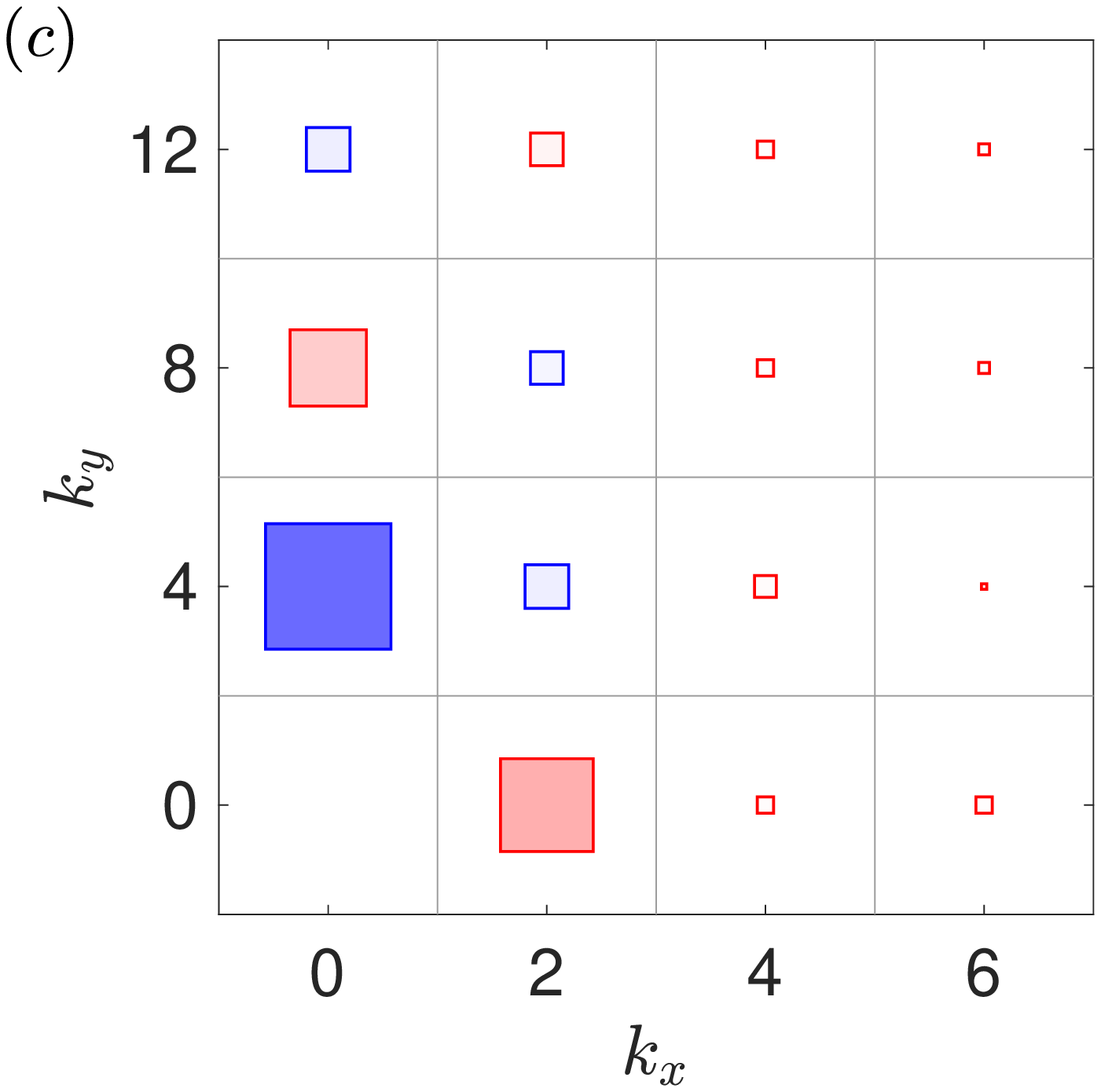}
	
	\caption{Contributions of (a) production, (b) dissipation and (c) nonlinear transfer to the energy balance of each Fourier mode for P4U. Both colour intensity and area of the square marker indicate each term's magnitude. The colours red and blue denote positive and negative quantities, respectively.}\label{fig:ecs balance}
\end{figure}

Production, dissipation and nonlinear transfer are computed for P4U and are illustrated in figure \ref{fig:ecs balance} for the subset of wavenumber pairs discussed in the previous section. A square marker appears at the centre of each tile. Both colour intensity and area of the square marker indicate each term's magnitude. The colours red and blue denote positive and negative quantities, respectively. In order to satisfy (\ref{eq:budget}), the sum across tiles which appear in the same position in each of the three figure panels must be zero. Additionally, the sum over all tiles in figure \ref{fig:ecs balance}(c) is approximately zero since the nonlinear terms are conservative when summed over all scales (this sum would be exactly zero if all wavenumber pairs were displayed in the figure).

Half of the production terms in figure \ref{fig:ecs balance}(a) are positive with the largest energy-producing modes being the streamwise-constant modes. The maximum production occurs at $(k_x,k_y) = (0,4)$, which is also the most energetic mode in the flow (see figure \ref{fig:energy}a). Production is negative for some scales. Of particular note is that production is negative for all of the spanwise-constant modes. Even though $\hat{P} \approx 0$ for most of these spanwise-constant modes, the same cannot be said for $(2,0)$ for which the production is negative and of large amplitude. In fact, its magnitude is comparable to that of $(0,8)$ even though it is less energetic, i.e. $|\hat{P}(2,0)| \approx |\hat{P}(0,8)|$ even though $\hat{E}(2,0) < \hat{E}(0,8)$. As expected, all dissipation terms in figure \ref{fig:ecs balance}(b) are negative. 

The nonlinear transfer in figure \ref{fig:ecs balance}(c) contains both positive and negative terms as the sum over all scales must equal zero. Consistent with the turbulent cascade, most values are positive, indicating that they are receiving energy from nonlinear transfer. The most notable exception is the $(0,4)$ mode, which must redistribute energy to other scales since dissipation offsets less than half of production. The additional scales that lose energy due to nonlinear transfer all have low streamwise wavenumbers. The (0,8) mode is one that receives energy from nonlinear transfer since $\hat{D}(0,8) > \hat{P}(0,8)$. Perhaps surprisingly, the spanwise-constant modes receive a considerable share of the nonlinearly-transferred energy. In particular the (2,0) mode receives more energy than any other mode. The (4,0) and (6,0) modes also receive rather than donate energy. Therefore, in addition to a cascade of energy from large scales to small scales, there is also a significant transfer from scales that are streamwise-constant to scales that are spanwise-constant. Indeed, the (2,0) mode (the largest recipient) is in fact larger in scale than the (0,4) mode (the largest donor). Thus in addition to a cascade, there also exists a transfer to scales of a similar scale but with a different orientation of their wavenumber vector.

\subsection{Resolvent predictions} \label{sec:P4 resolvent}

\begin{figure}
	\centering
	\includegraphics[trim = 1.8cm 0cm 3.2cm 0cm, clip,scale=0.3]{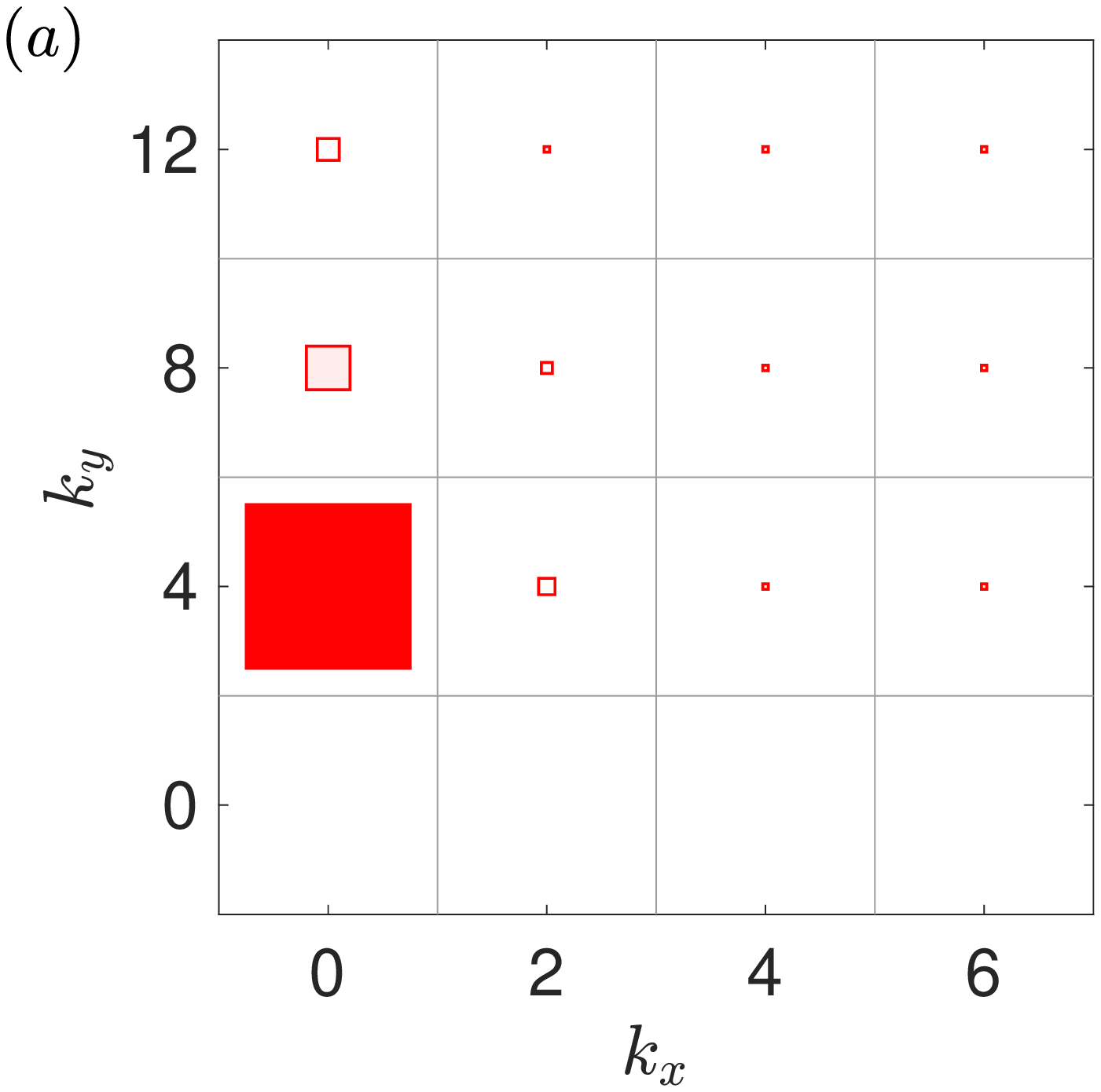}
	\includegraphics[trim = 1.8cm 0cm 3.2cm 0cm, clip,scale=0.3]{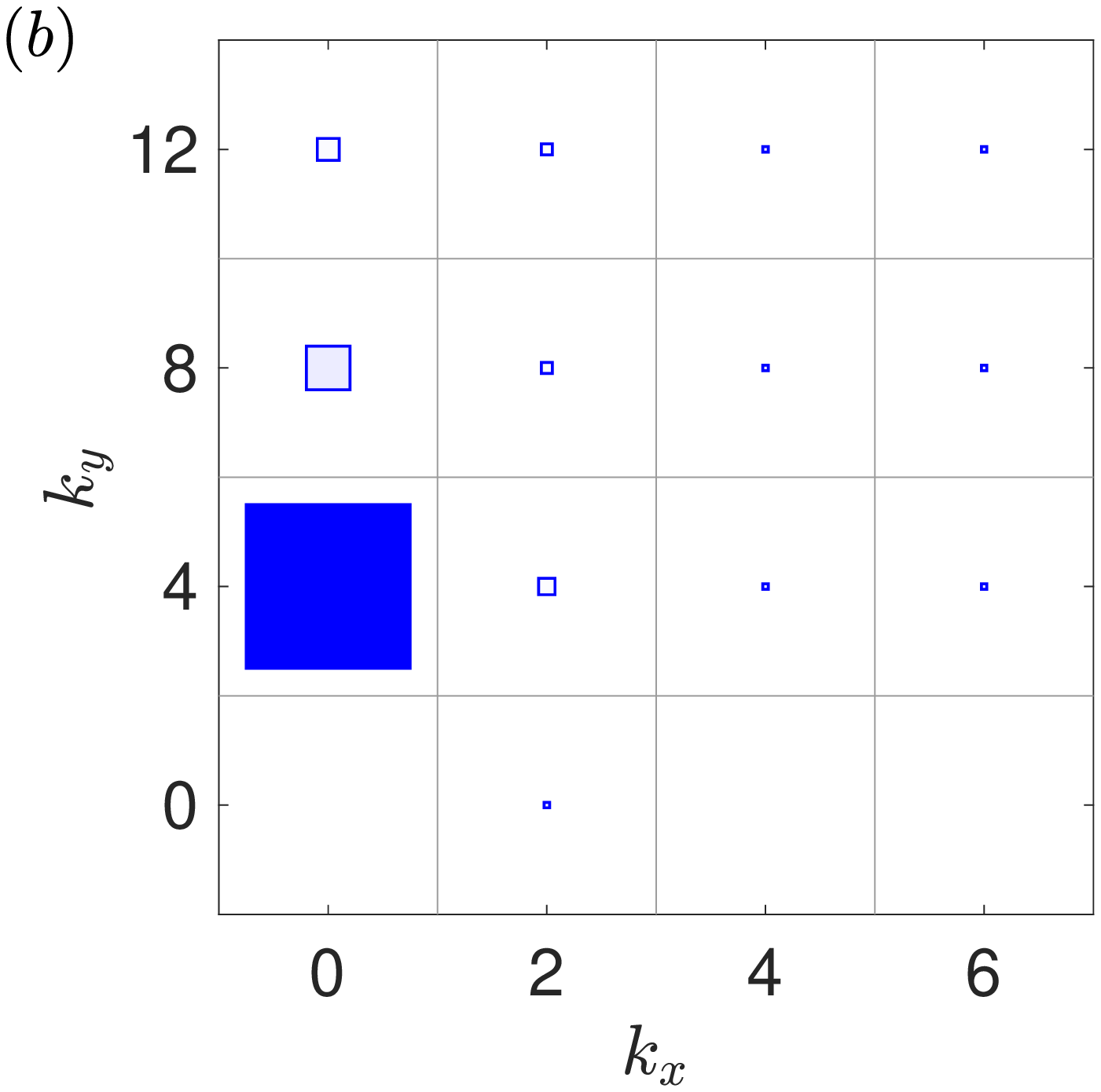}
	\includegraphics[trim = 1.8cm 0cm 3.2cm 0cm, clip,scale=0.3]{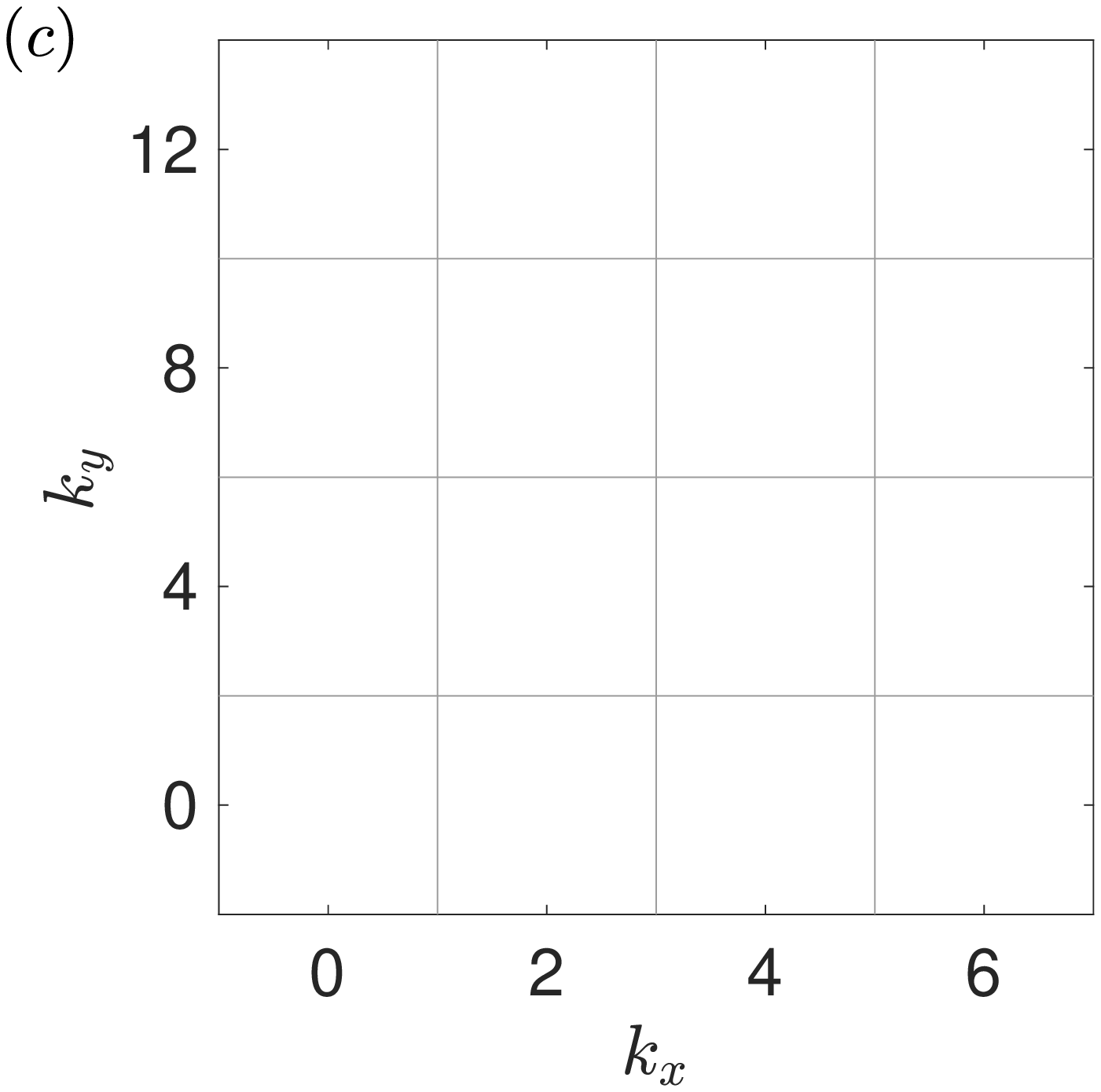}
	
	\caption{Contributions of (a) production, (b) dissipation and (c) nonlinear transfer from the first P4U resolvent mode.}\label{fig:ecs resolvent 1}
\end{figure}

Having considered the true energy balance from (\ref{eq:budget}), we now focus on its counterpart for the first resolvent mode in (\ref{eq:budget rank 1}). To do so, it is necessary to set $\omega = c^+k_x$ since the wave speed is fixed at $c^+ = 14.2$. Figure \ref{fig:ecs resolvent 1} illustrates the production, dissipation and nonlinear transfer in a manner analogous to that of figure \ref{fig:ecs balance}. The resolvent prediction for production in figure \ref{fig:ecs resolvent 1}(a) is positive for every scale and the largest value occurs when $(k_x,k_y) = (0,4)$. The predictions for the largest scales are similar to the true values of production in figure \ref{fig:ecs balance}(a) and reflect the resolvent operator's ability to identify linear amplification mechanisms. 

The  dissipation and nonlinear transfer from the first resolvent mode in figures \ref{fig:ecs resolvent 1}(b) and (c), respectively, are less similar to the true values in figures \ref{fig:ecs balance}(b) and (c). For all scales, the dissipation is nearly equal and opposite to production resulting in very small values for nonlinear transfer. A similar phenomenon is observed by \cite{Jin20} for the first resolvent mode in low Reynolds number cylinder flow. It can therefore be concluded that suboptimal resolvent modes are necessary to correctly model nonlinear transfer.

\subsection{Eddy dissipation} \label{sec:P4 eddy}

\begin{figure}
	\centering
	\includegraphics[trim = 1.8cm 0cm 3.2cm 0cm, clip,scale=0.3]{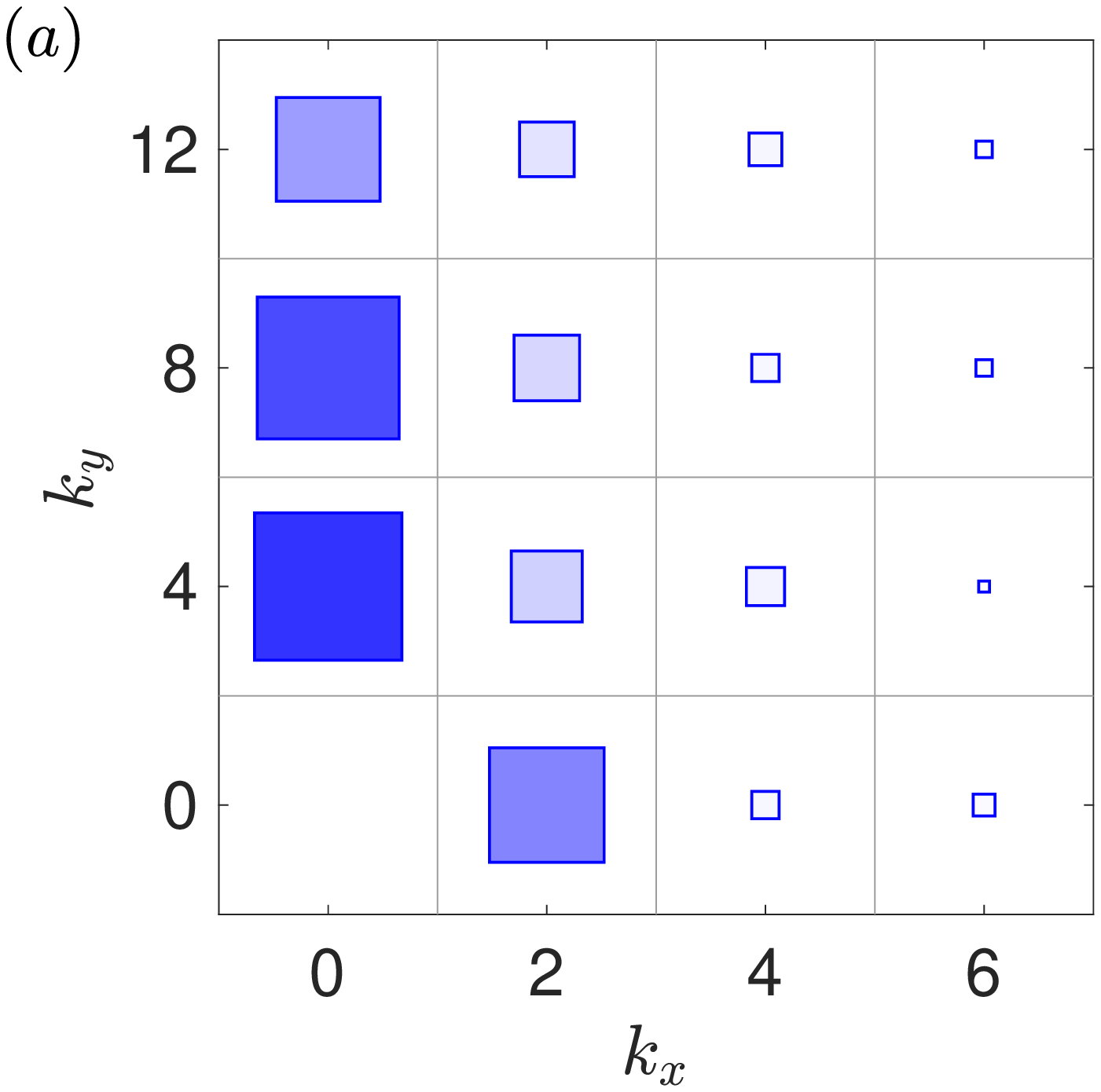}
	\includegraphics[trim = 1cm 0cm 2.5cm 0cm, clip,scale=0.3]{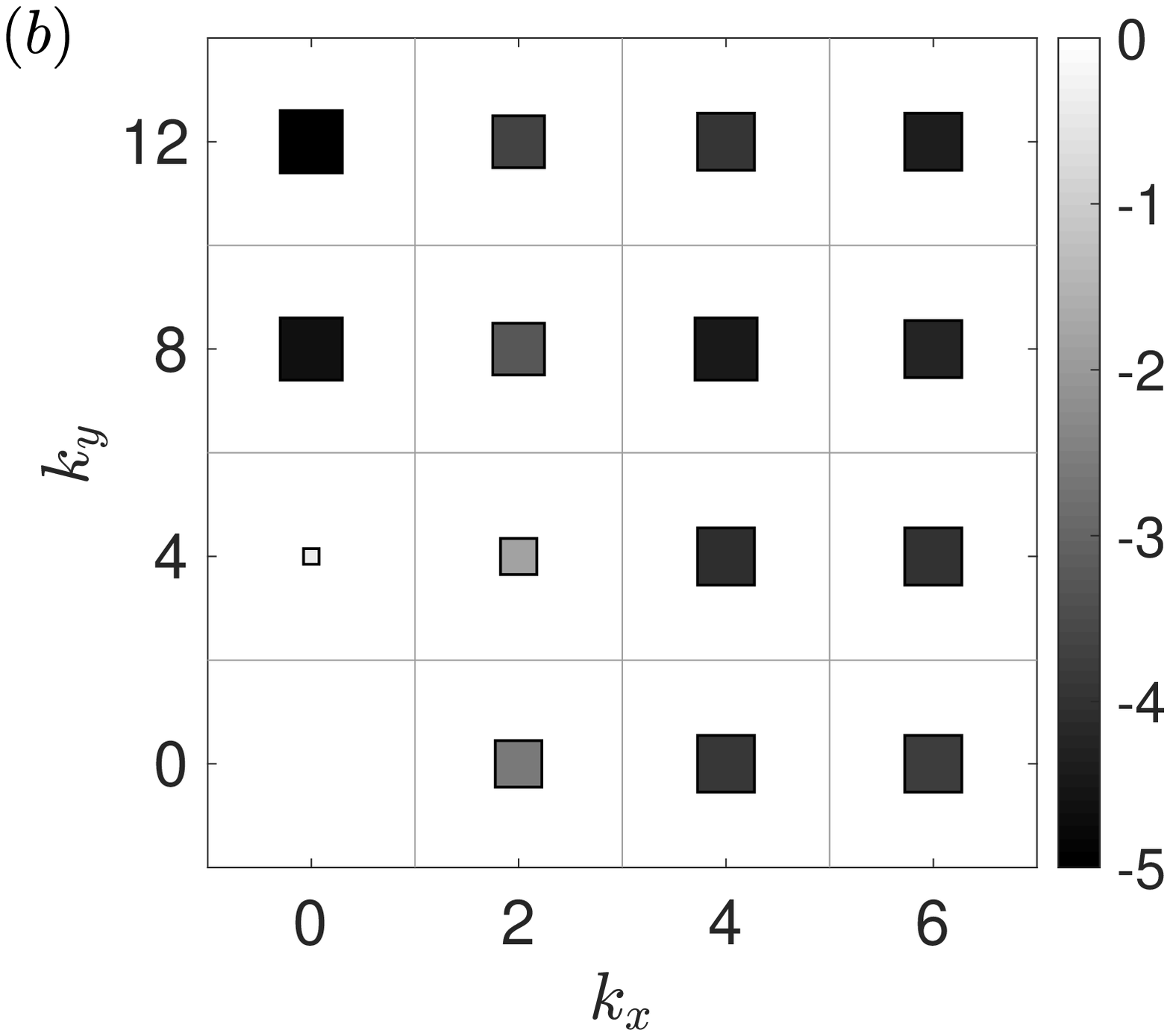}
	
	\caption{(a) Nonlinear transfer modelled by eddy viscosity for P4U and (b) its error compared to the true nonlinear transfer in figure \ref{fig:ecs balance}(c).}\label{fig:ecs eddy}
\end{figure}

As discussed in \S\ref{sec:eddy viscosity}, one way to model nonlinear transfer is through the use of an eddy viscosity. Figure \ref{fig:ecs eddy}(a) presents the eddy dissipation from (\ref{eq:eddy dissipation}), which is negative for all wavenumber pairs considered. Unlike nonlinear transfer, therefore, eddy dissipation is not conservative and contributes net energy loss to every scale. Ideally the eddy dissipation would resemble $\hat{N}(k_x,k_y)$ in figure \ref{fig:ecs balance}(c), so its error $\epsilon(k_x,k_y)$ with respect to nonlinear transfer is computed in figure \ref{fig:ecs eddy}(b) using the expression
\begin{equation} \label{eq:epsilon}
\epsilon(k_x,k_y) = \frac{\widehat{Edd}(k_x,k_y) - \hat{N}(k_x,k_y)}{|\hat{N}(k_x,k_y)|}.
\end{equation}
The error for all wavenumber pairs exceeds 1 other than $(0,4)$ where $\epsilon \approx 0.28$. The size of the square marker in figure \ref{fig:ecs eddy}(b) reflects the magnitude of the error and the smallest marker coincides with the tile belonging to $(0,4)$. The fact that $\epsilon$ is lowest for this scale indicates that the eddy viscosity is most effective for highly amplified linear mechanisms. In other words, the eddy viscosity works best for scales where viscous dissipation is not sufficient to balance production.

\section{Results: Minimal channel} \label{sec:channel}

Having analysed the energy transfer for the P4U ECS, this section examines the same quantities for the minimal channel. Since each wavenumber pair has a distribution of temporal frequencies, all terms in the energy balance are time-averaged. 

\subsection{DNS and resolvent energy balances}

\begin{figure}
	\centering
	\includegraphics[trim = 1.8cm 0cm 3.2cm 0cm, clip,scale=0.3]{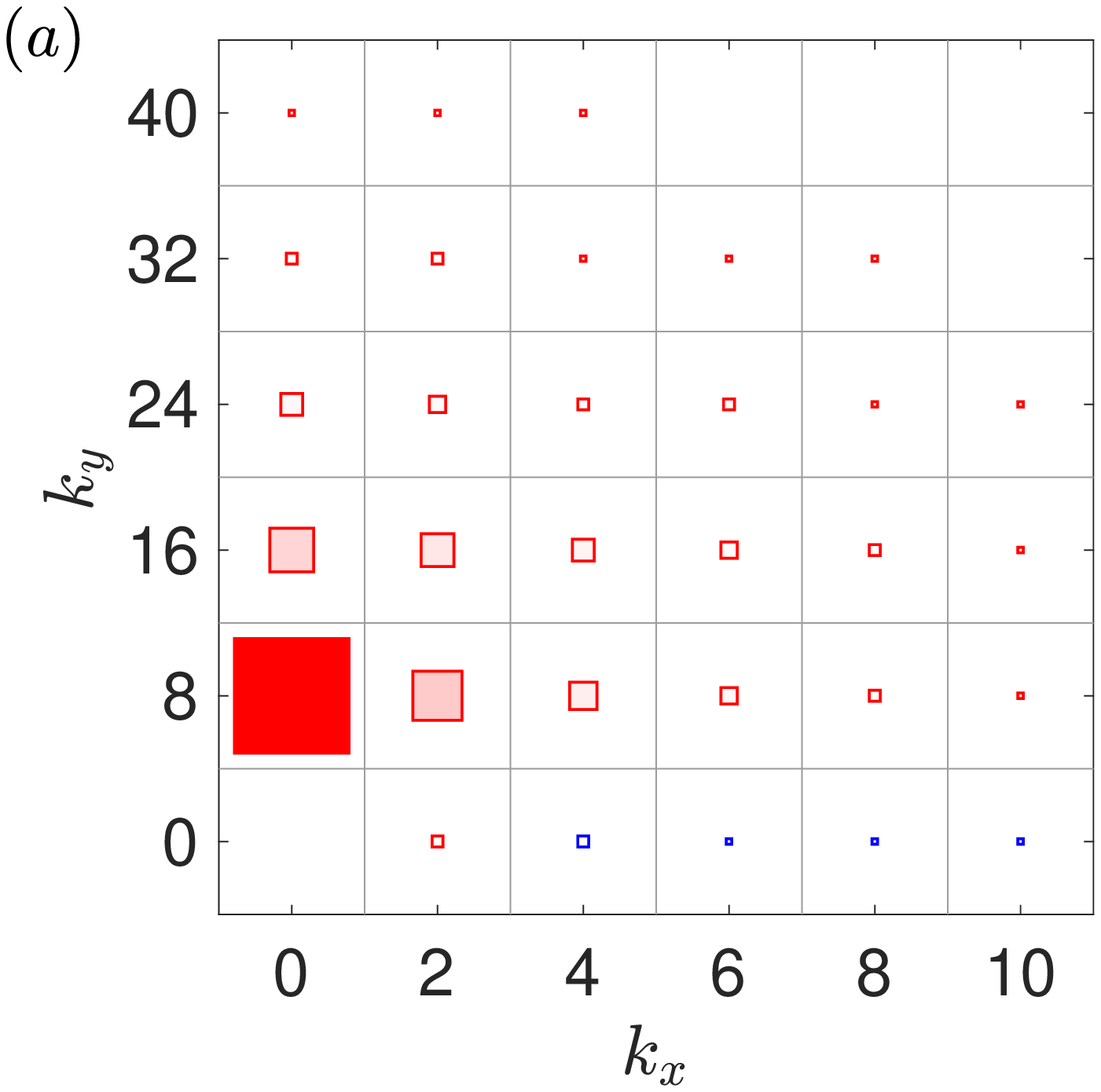}
	\includegraphics[trim = 1.8cm 0cm 3.2cm 0cm, clip,scale=0.3]{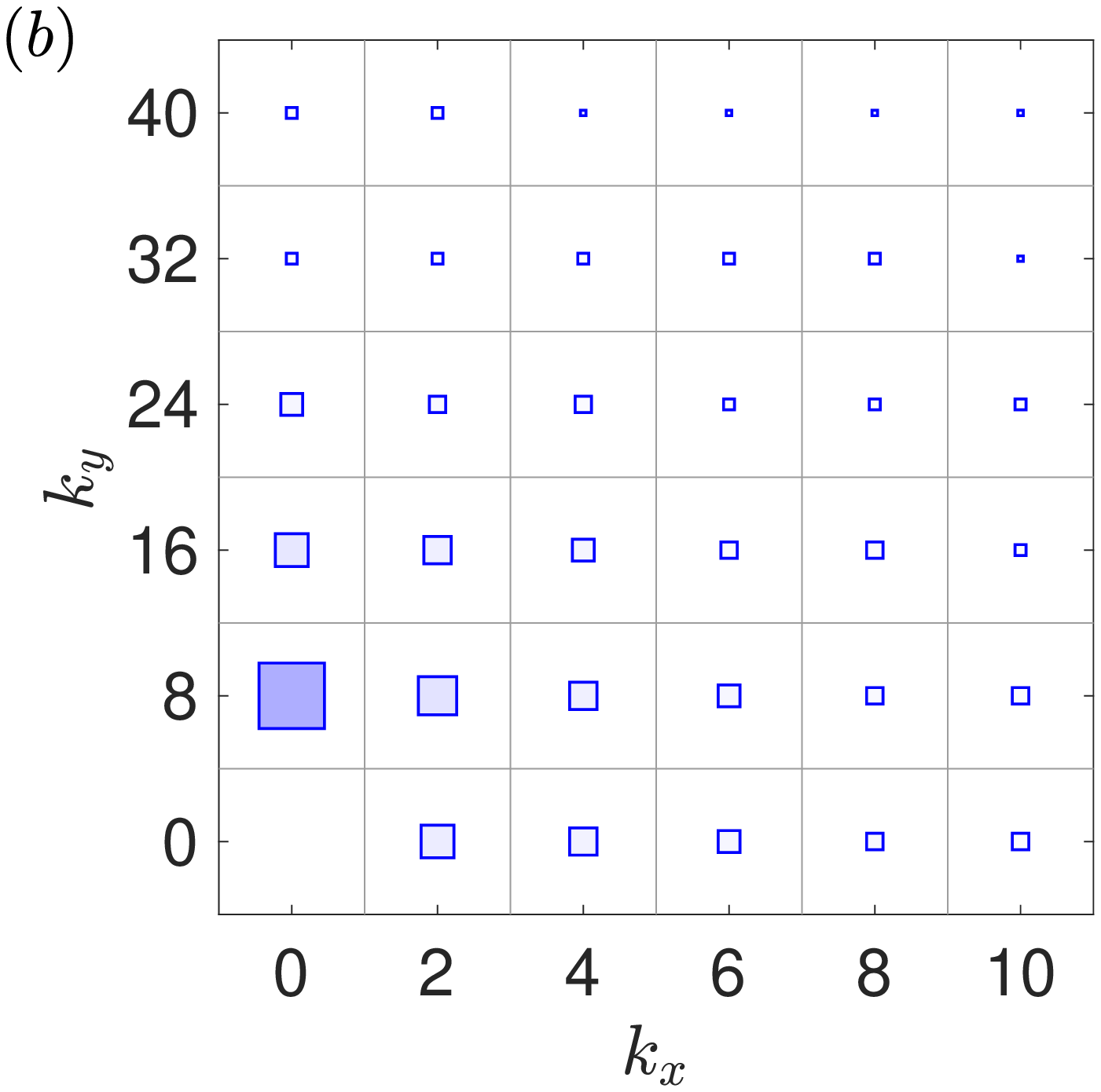}
	\includegraphics[trim = 1.8cm 0cm 3.2cm 0cm, clip,scale=0.3]{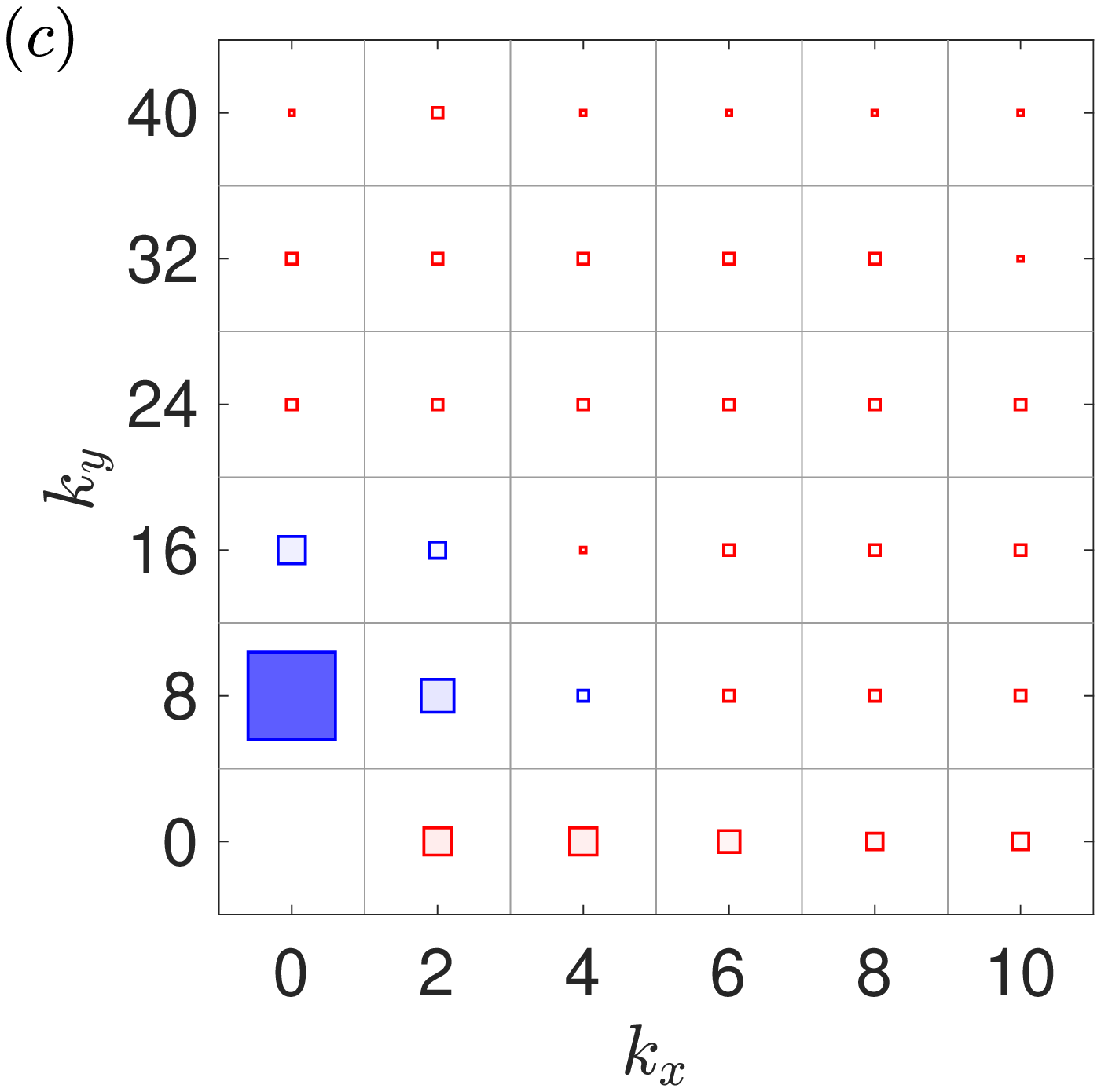}
	
	\caption{Contributions of (a) production, (b) dissipation and (c) nonlinear transfer to the energy balance of each Fourier mode for the minimal channel.}\label{fig:minchan balance}
\end{figure}

Production, dissipation and nonlinear transfer for the minimal channel are illustrated in figure \ref{fig:minchan balance}. For almost all wavenumber pairs shown, production is positive as seen in figure \ref{fig:minchan balance}(a) with the maximum occurring for $(0,8)$. The only scales where production is negative are spanwise-constant, i.e. $k_y = 0$. The dissipation in figure \ref{fig:minchan balance}(b) is negative for all scales, as expected. Even though the largest dissipation occurs for $(0,8)$, its value is comparable to that for other wavenumbers. 

The nonlinear transfer in figure \ref{fig:minchan balance}(c) illustrates that the surfeit of energy not dissipated by viscosity from $(0,8)$ is redistributed to other scales. Moreover, all scales which lose energy due to nonlinear transfer are clustered around low streamwise wavenumbers. This is consistent with the turbulence cascade in that energy from the large-scales trickles down to smaller scales which are more effective at dissipating energy. All but five scales in figure \ref{fig:minchan balance}(c) receive energy from nonlinear transfer with the largest amounts going to spanwise-constant structures. This is interesting given that these Tollmien-Schlichting-type waves are the first to become unstable \citep{Tollmien29,Schlichting33} yet they play a damping role in the minimal channel. 

\begin{figure}
	\centering
	\includegraphics[trim = 1.8cm 0cm 3.2cm 0cm, clip,scale=0.3]{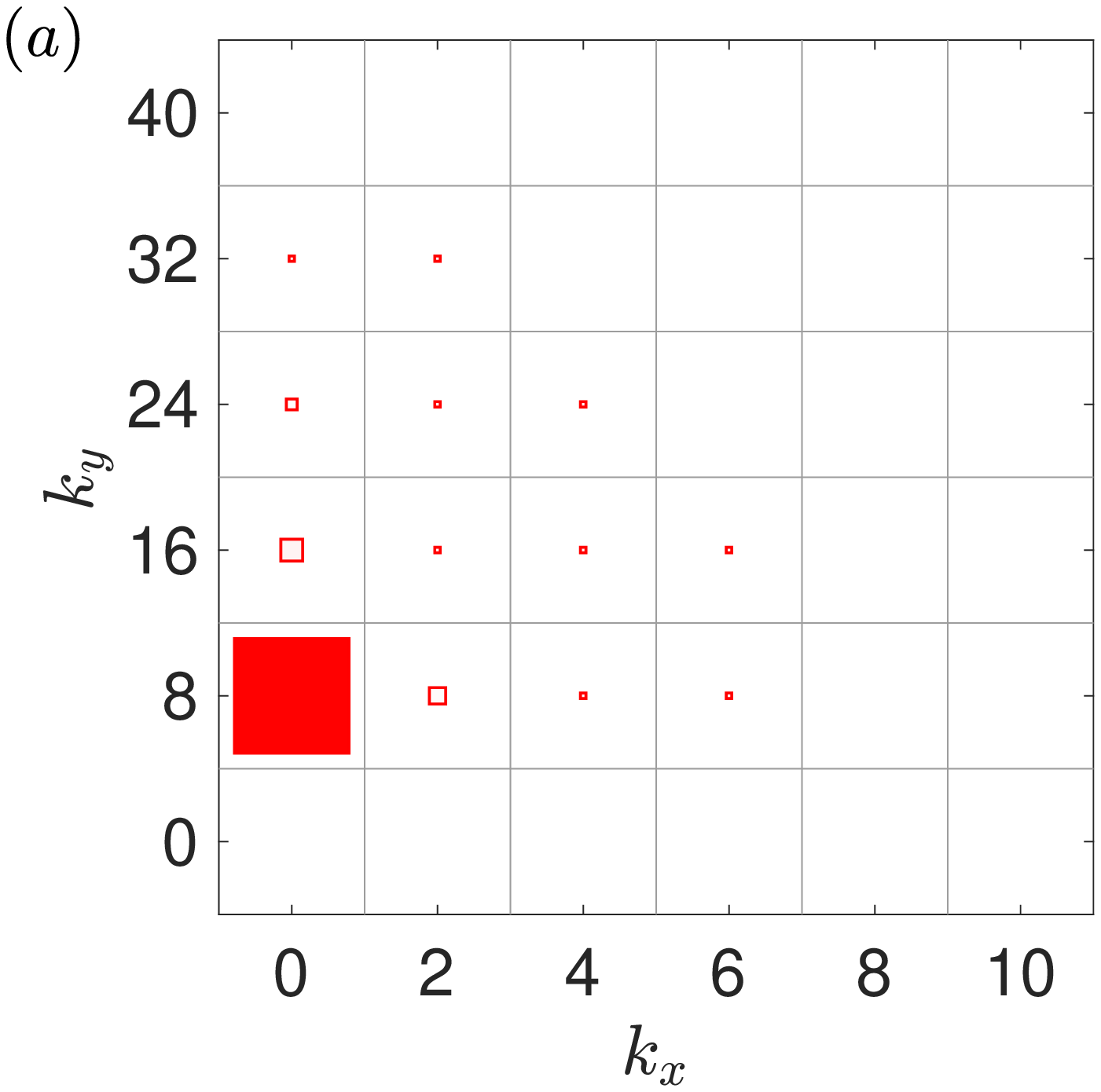}
	\includegraphics[trim = 1.8cm 0cm 3.2cm 0cm, clip,scale=0.3]{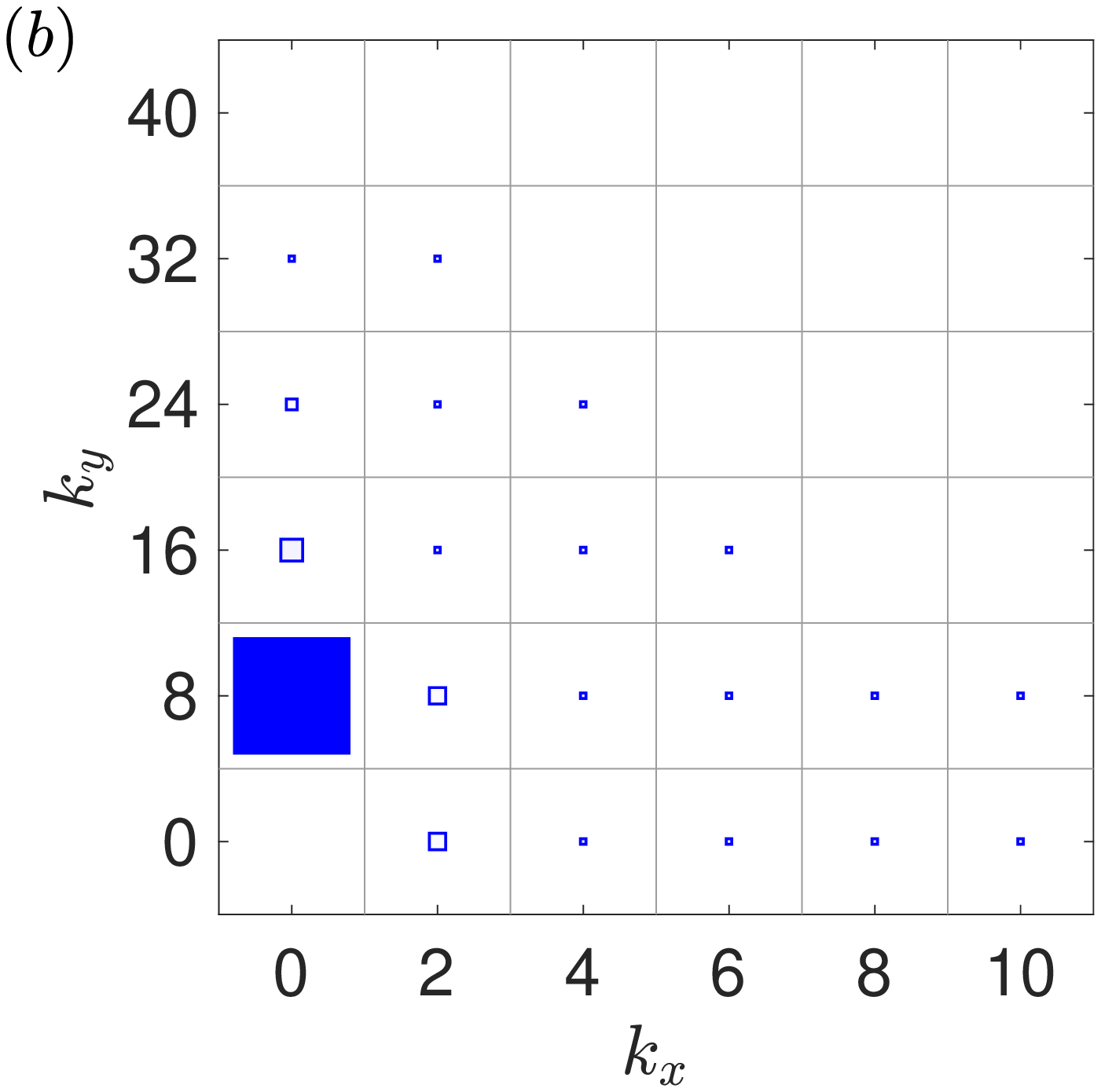}
	\includegraphics[trim = 1.8cm 0cm 3.2cm 0cm, clip,scale=0.3]{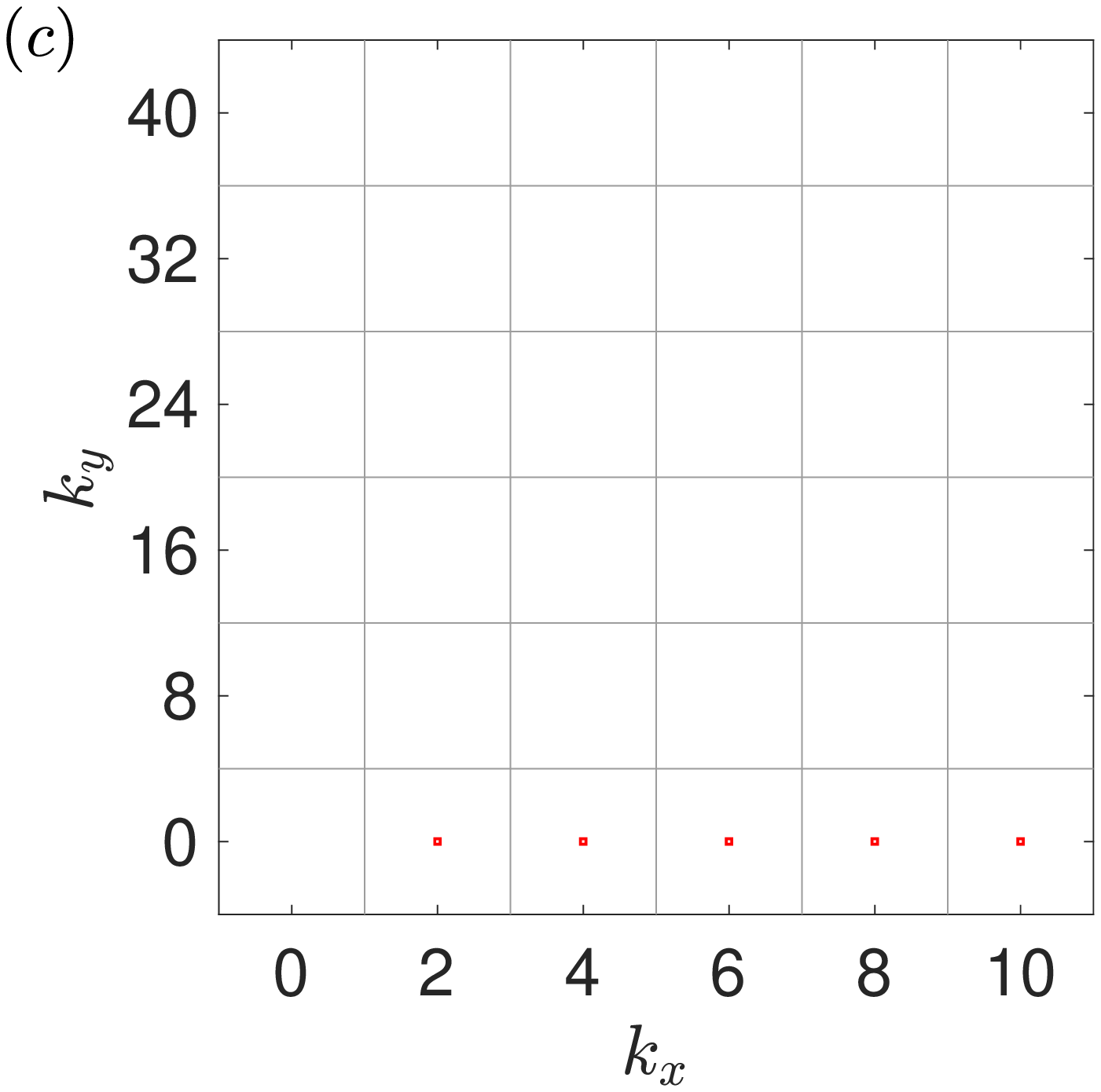}
	
	\caption{Contributions of (a) production, (b) dissipation and (c) nonlinear transfer from the first resolvent mode in the case of the minimal channel.}\label{fig:minchan resolvent 1}
\end{figure}

The energy balance for the first resolvent mode is presented in figure \ref{fig:minchan resolvent 1}. Since each wavenumber pair has a distribution of energetic temporal frequencies, we compute the singular values across a discretisation of $\omega$ and choose the $\omega$ that results in the largest amplification. As one example, $\omega = 0$ leads to the largest amplification for $(0,8)$. The production in figure \ref{fig:minchan resolvent 1} is so large for this scale that the choice of $\omega$ for other scales has little impact on the result. It can be remarked that viscous dissipation in figure \ref{fig:minchan resolvent 1}(b) is sufficient to completely counteract production for the majority of scales considered. Since the sum of all three terms must be zero for each resolvent mode, it follows that nonlinear transfer is negligible for nearly all scales as seen in figure \ref{fig:minchan resolvent 1}(c). 

\begin{figure}
	\centering
	\includegraphics[trim = 1.8cm 0cm 3.2cm 0cm, clip,scale=0.3]{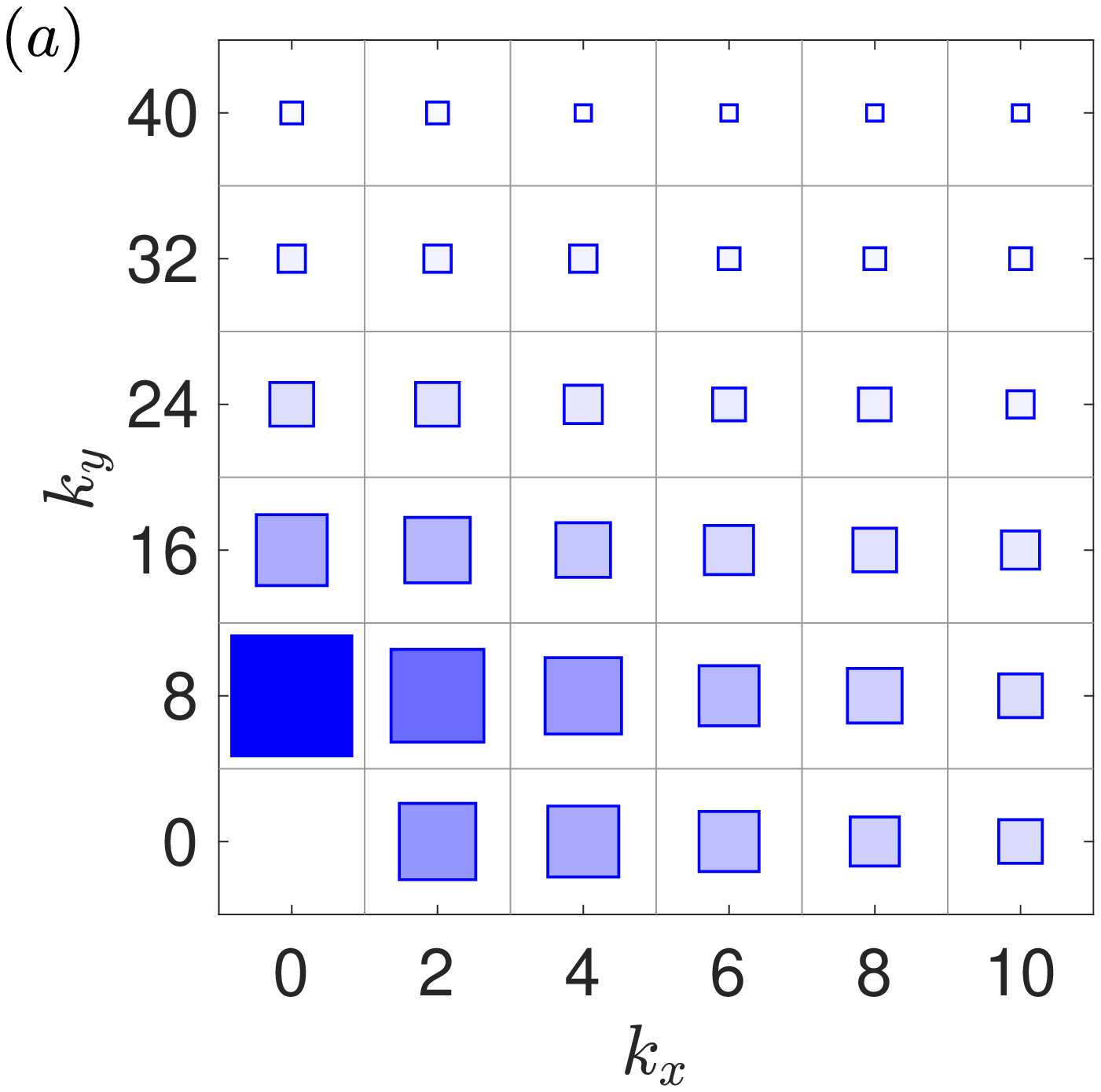}
	\includegraphics[trim = 1.0cm 0cm 2.0cm 0cm, clip,scale=0.3]{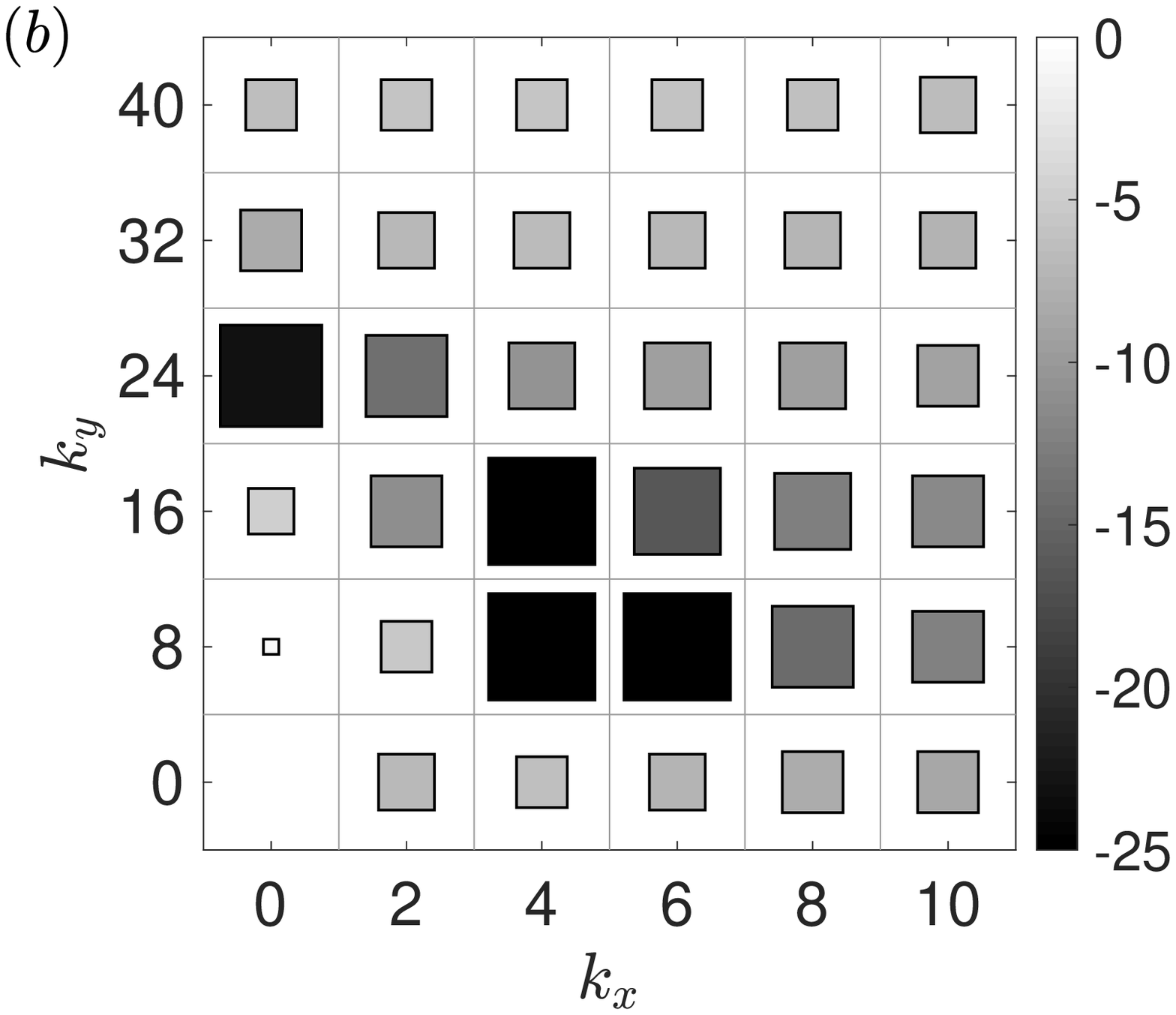}
	
	\caption{(a) Nonlinear transfer modelled by eddy viscosity for the minimal channel and (b) its error compared to the true nonlinear transfer in figure \ref{fig:minchan balance}(c).}\label{fig:minchan eddy}
\end{figure}

To identify for which scales eddy viscosity can model nonlinear transfer, the eddy dissipation is computed and displayed in figure \ref{fig:minchan eddy}(a). As expected, it is negative for all scales even though nonlinear transfer tends to be positive outside the cluster around $(0,8)$. The error $\epsilon$, as defined in (\ref{eq:epsilon}), is thus large for the majority of scales as seen in figure \ref{fig:minchan eddy}(b). The only scale where $\epsilon < 1$ is $(0,8)$. Although $\epsilon > 1$ for every other scale, those where nonlinear transfer is negative such as $(0,16)$ or $(2,8)$ have lower values of $\epsilon$ than scales where nonlinear transfer is positive. 

\subsection{Comparison of P4U and minimal channel}

The energy transfer processes in P4U are similar to those of the minimal channel. Production is positive for the majority of scales and its maximum occurs for $(\lambda_x^+,\lambda_y^+) \approx (\infty,100)$, which corresponds to $(k_x,k_y) = (0,4)$ in P4U and $(k_x,k_y) = (0,8)$ in the minimal channel. The production for spanwise-constant scales, on the other hand, is mostly negative. Dissipation is always negative but it is insufficiently large to counterbalance production for the largest streamwise-constant scales. Nonlinear transfer contains both positive and negative terms as the sum over all scales must be zero according to (\ref{eq:conservative}). Energy is primarily removed from the largest structures and redistributed to smaller ones. Interestingly, there is also a polarisation effect where nonlinear transfer reallocates energy from streamwise-constant to spanwise-constant modes. Finally, our choice to integrate over the wall-normal domain and analyse energy transfer in $(k_x,k_y)$ space permits quantification of energy loss due to eddy dissipation. For both flows, the only scale where eddy dissipation can quantitatively predict energy loss due to nonlinear transfer is the most energetic scale $(\lambda_x^+,\lambda_y^+) \approx (\infty,100)$. 

The only notable difference between the two flows is that nonlinear transfer has a clearer pattern for the minimal channel. Scales where this term is negative are localised in one cluster of large structures. For P4U, there is more scatter primarily due to the $(0,8)$ mode which is able to dissipate more energy than it produces. 

\section{Non-normality in the energy balance} \label{sec:discussion}

The results in \S\S\ref{sec:ECS} and \ref{sec:channel} indicate that the optimal resolvent mode does not accurately account for energy transfer between scales. As we will explain in \S\ref{sec:nonnormality}, the root of this discrepancy is non-normality induced by the mean shear. In \S\ref{sec:counteract}, we show that its influence can be weakened by eddy viscosity. The efficiency of resolvent modes, therefore, is lower than that for eddy modes when reconstructing the energy balance in \S\ref{sec:projection} using correctly weighted resolvent and eddy modes. Finally, we highlight in \S\ref{sec:aspect ratio} that the eddy viscosity is most effective for high aspect ratio modes where the influence of non-normality is most pronounced. 

\subsection{Competition between production and nonlinear transfer} \label{sec:nonnormality}

In this section, we demonstrate that non-normality leads to a competition between the production and nonlinear transfer terms in the energy balance. To simplify the discussion, we will only consider the first resolvent mode although similar arguments can be made for suboptimal modes. We begin by rewriting the resolvent norm, or the first singular value $\sigma_1(\boldsymbol{k})$, as
\begin{equation} \label{eq:sigma1}
\sigma_1(\boldsymbol{k}) \approx \frac{1}{(i\omega - \lambda_{min}(\boldsymbol{k}))} \cdot \frac{1}{\int_{-h}^{h} \hat{\boldsymbol{\psi}}_1^*(\boldsymbol{k})\hat{\boldsymbol{\phi}}_1(\boldsymbol{k})dz}.
\end{equation}
Equation (\ref{eq:sigma1}) is a result from \cite{Symon18} and is applicable for resonant mechanisms such as $k_x = 0$ modes in channel flow or the shedding mode in cylinder flow. In (\ref{eq:sigma1}), the resolvent norm is rewritten as the product of two terms. The first is the inverse distance between the imaginary axis and the least stable eigenvalue $\lambda_{min}(\boldsymbol{k})$ of the linear Navier-Stokes operator $\boldsymbol{A}$. The second is a metric of non-normality, originally proposed by \cite{Chomaz05}, which is equivalent to the inverse of nonlinear transfer in (\ref{eq:res mode balance}). In parallel shear flows, the lift-up mechanism \citep{Landahl80} results in component-type non-normality \citep{Marquet09} so all the energy for $\hat{\boldsymbol{\psi}}_1$ is concentrated in $\hat{u}$ while all the energy for $\hat{\boldsymbol{\phi}}_1$ is concentrated in $\hat{v}$ and $\hat{w}$. Consequently, non-normality results in $\sigma_1$ being very large.

We proceed by designating $C_1 = (i\omega - \lambda_{min})$ and substituting (\ref{eq:sigma1}) into (\ref{eq:budget rank 1}) to arrive at the following 
\begin{equation} \label{eq:non-normality}
\frac{1}{C_1 \int_{-h}^{h} \hat{\boldsymbol{\psi}}_1^*(\boldsymbol{k})\hat{\boldsymbol{\phi}}_1(\boldsymbol{k})}  \left( \check{P}(\boldsymbol{k})) + \check{D}(\boldsymbol{k}) \right) + 
\underbrace{\int_{-h}^h \boldsymbol{\psi}_1^*(\boldsymbol{k}) \hat{\boldsymbol{\phi}}_1(\boldsymbol{k})dz}_{\hat{N}(\boldsymbol{k})} = 0,
\end{equation}
where $\check{(\cdot)}$ denotes a quantity that has been normalised by $\sigma_1(\boldsymbol{k})$. Equation (\ref{eq:non-normality}) illustrates that as non-normality increases (thus leading to higher amplification), so does the disparity between production and nonlinear transfer. This behaviour is problematic because even for the low Reynolds number flows considered here, we find in the DNS that dissipation is not sufficient to counteract production. Therefore, the nonlinear transfer term cannot be small because it needs to remove a considerable amount of energy for the scale to reach equilibrium. Despite identifying amplification mechanisms, non-normality therefore hinders the efficiency of the optimal resolvent mode in representing the true velocity fluctuations by hindering energy transfer between scales.

As an aside, we note that the same arguments are applicable to cylinder flow for which \cite{Jin20} found that nonlinear transfer for the first resolvent mode was nearly zero. The primary difference for spatially-developing flows is that convective non-normality \citep{Chomaz05, Marquet09, Symon18} is responsible for minimising $\hat{\boldsymbol{\psi}}_1^* \hat{\boldsymbol{\phi}}_1$. Mean advection in the resolvent operator localises $\hat{\boldsymbol{\psi}}_1$ downstream of the cylinder and $\hat{\boldsymbol{\phi}}_1$ upstream of the cylinder. The non-normality increases amplification by at least one order of magnitude \citep{Symon18} but leads to a drastic underestimation of nonlinear transfer by the first resolvent mode. 

\subsection{Counteracting non-normality with eddy viscosity} \label{sec:counteract}

\begin{figure}
	\centering
	\includegraphics[trim = 0.1cm 0cm 1.5cm 0cm, clip,scale=0.24]{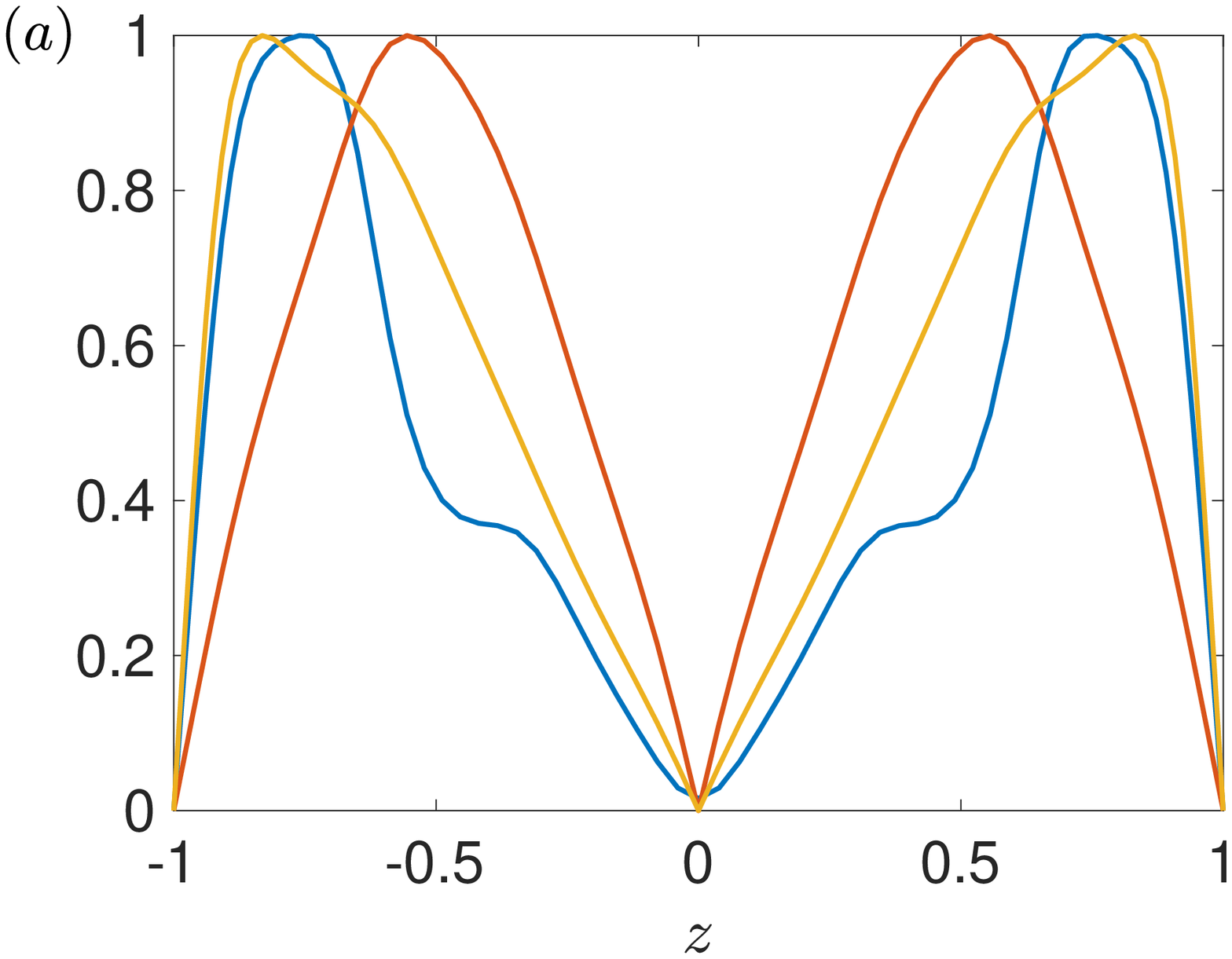}
	\includegraphics[trim = 0.1cm 0cm 1.5cm 0cm, clip,scale=0.24]{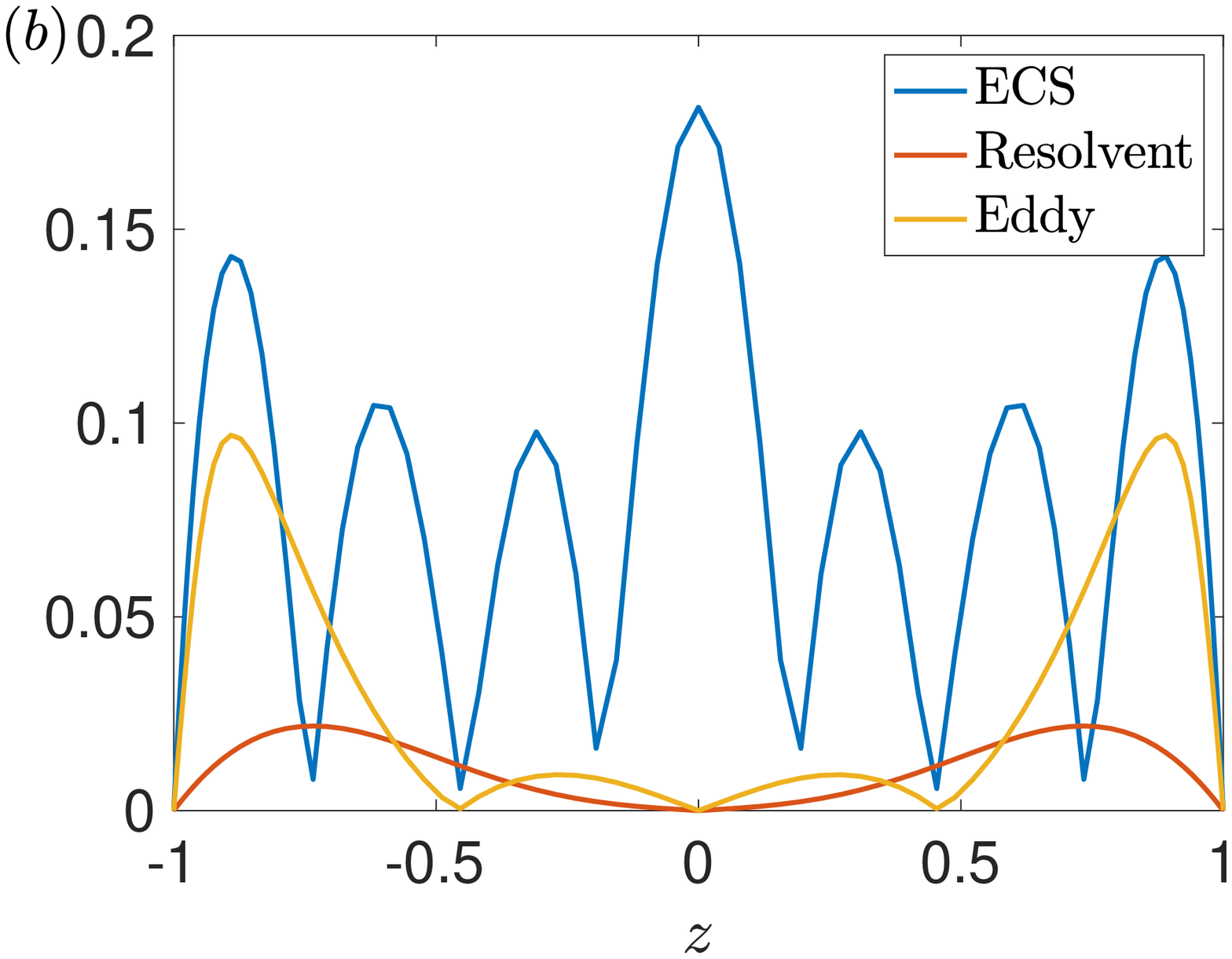}
	\includegraphics[trim = 0.1cm 0cm 1.5cm 0cm, clip,scale=0.24]{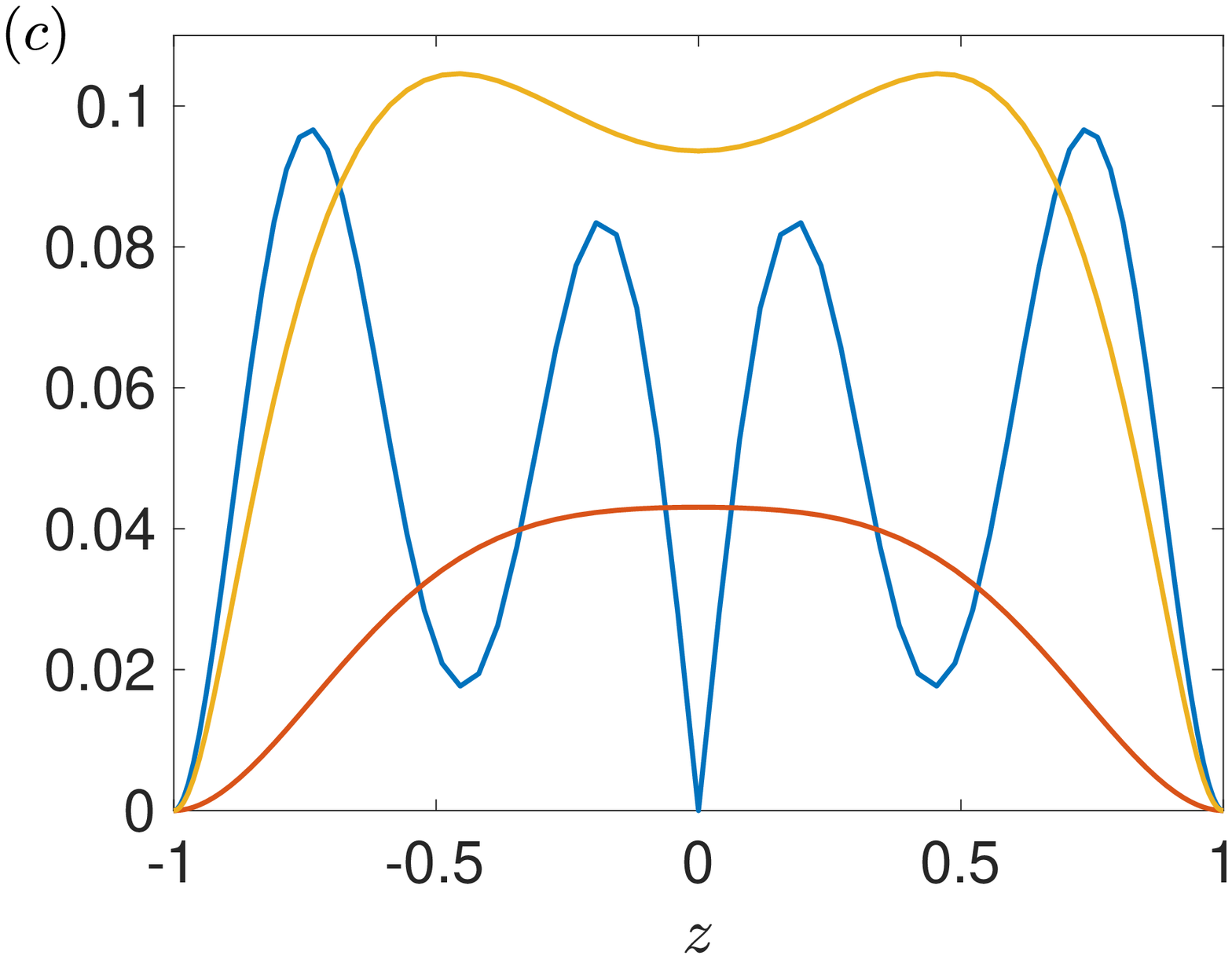}
	
	\caption{The (a) streamwise, (b) spanwise and (c) wall-normal velocity components of $(k_x,k_y) = (0,4)$. The first resolvent mode is in blue, the first eddy mode is in orange, and the DNS is in yellow. The modes are normalised by the peak value of the streamwise velocity component.}\label{fig:mode comparison}
\end{figure}

Returning to the channel flow case, the root of non-normality is the coupling term $-ik_yU'$ of the resolvent operator. Since only the spanwise wavenumber appears in this term, higher aspect ratio structures where $k_x < k_y$ are more prone to amplification. The simplest way to explain the weakening of non-normality by eddy viscosity is that it damps the linear operator. In other words, the effect of viscosity is increased to counteract the mean shear. As a representative example, we plot the first resolvent and eddy mode in figure \ref{fig:mode comparison} for $(k_x,k_y) = (0,4)$ and compare it to the DNS, i.e. the true Fourier mode for P4U.  
In terms of the streamwise velocity component, the first eddy mode is in better agreement with the DNS than the first resolvent mode. In terms of the spanwise and wall-normal components, however, neither the eddy nor the resolvent mode are in close agreement with the DNS. Nevertheless, the eddy mode has more energy in these velocity components than its resolvent counterpart. Despite having a more complicated structure, the DNS also has more energy in $v$ and $w$, suggesting that the eddy viscosity has sufficiently dampened non-normality to provide a better basis for the flow. 

It is worth mentioning that eddy viscosity is not the only way to counteract mean shear. In \cite{Rosenberg19}, a componentwise analysis of the resolvent operator yielded two distinct families of modes which, when correctly weighted, destructively interfere to reduce bias towards the streamwise velocity component. The approach has been applied in \cite{McMullen20} to higher Reynolds number flows where the destructive interference is more pronounced due to stronger non-normality and higher mean shear. 

\subsection{Resolvent and eddy reconstructions of the energy budget} \label{sec:projection}

Since the first resolvent mode cannot provide the nonlinear transfer that was observed in the DNS, we hypothesise that the role of suboptimal modes in energy transfer is important. We test this hypothesis by reconstructing the energy budget from resolvent modes for P4U since there is a unique wave speed for every wavenumber pair (see appendix \ref{sec:SPOD} for the minimal channel case). The weights of resolvent modes $\tilde{\chi}_p(\boldsymbol{k})$ are determined by projecting them onto the velocity field as done in \cite{Sharma16}
\begin{equation} \label{eq:projection}
\tilde{\chi}_p(\boldsymbol{k}) = \sigma_p(\boldsymbol{k}) \chi_p (\boldsymbol{k})= \hat{\boldsymbol{\psi}}^*_p (\boldsymbol{k}) \hat{\boldsymbol{u}}(\boldsymbol{k}).
\end{equation}
The approximate velocity field $\hat{\boldsymbol{u}}_a(\boldsymbol{k})$ can be written as
\begin{equation} \label{eq:weights}
\hat{\boldsymbol{u}}_a(\boldsymbol{k}) = \sum_{p=1}^n \tilde{\chi}_p(\boldsymbol{k}) \hat{\boldsymbol{\psi}}_p(\boldsymbol{k}), 
\end{equation}
where $n$ is the number of resolvent modes used in the approximation. The reconstructed energy budget is evaluated as a function of $n$ by replacing $\hat{\boldsymbol{u}}(\boldsymbol{k})$ with $\hat{\boldsymbol{u}}_a(\boldsymbol{k})$ in (\ref{eq:budget k}).
Since eddy viscosity can lead to improvements, we will also do this for eddy modes by replacing $\hat{\boldsymbol{\psi}}_p(\boldsymbol{k})$ with $\boldsymbol{\psi}^e_p(\boldsymbol{k})$ in (\ref{eq:weights}). 

\begin{figure}
	\centering
	\includegraphics[trim = 0.1cm 0cm 1.5cm 0cm, clip,scale=0.36]{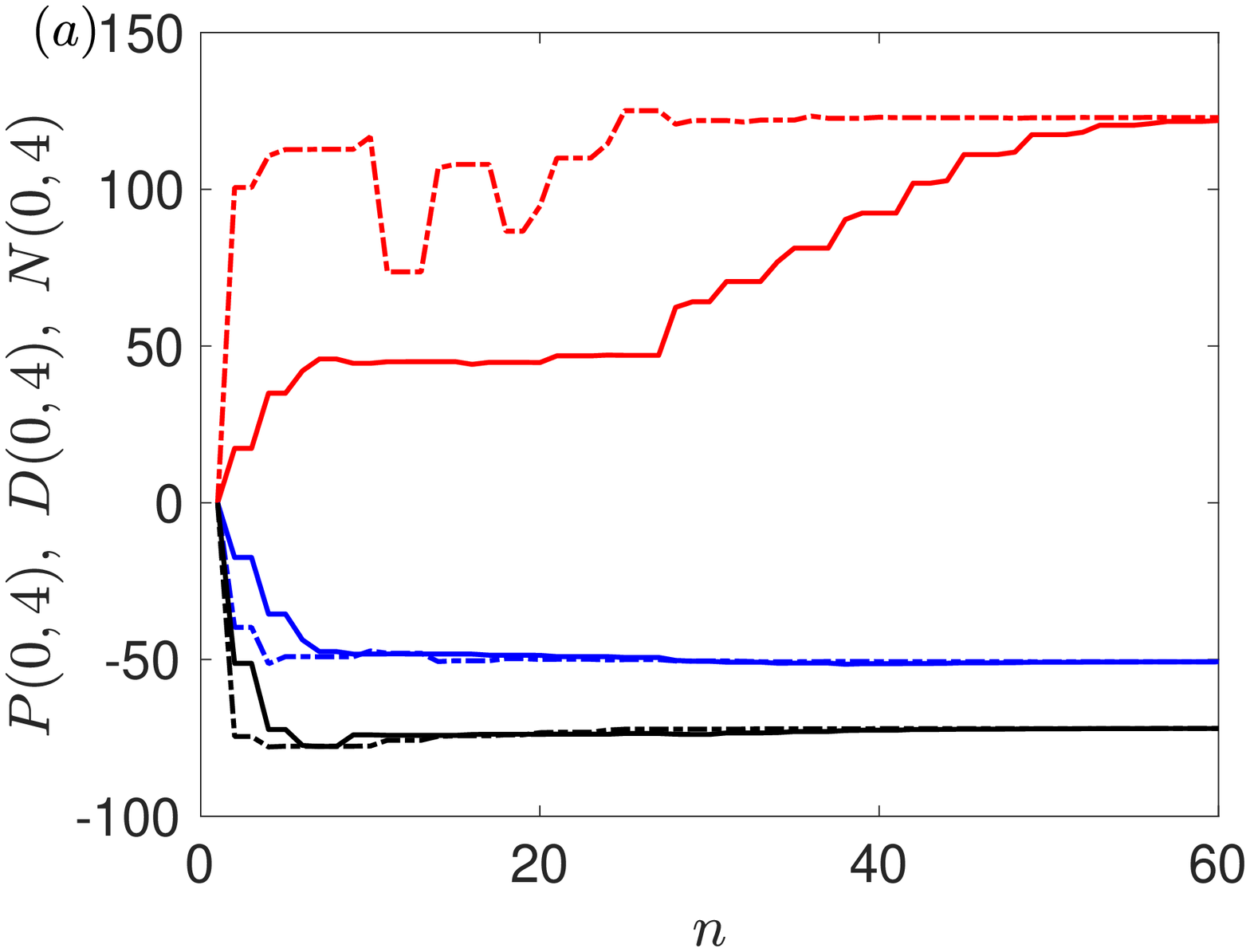}
	\includegraphics[trim = 0.1cm 0cm 1.5cm 0cm, clip,scale=0.36]{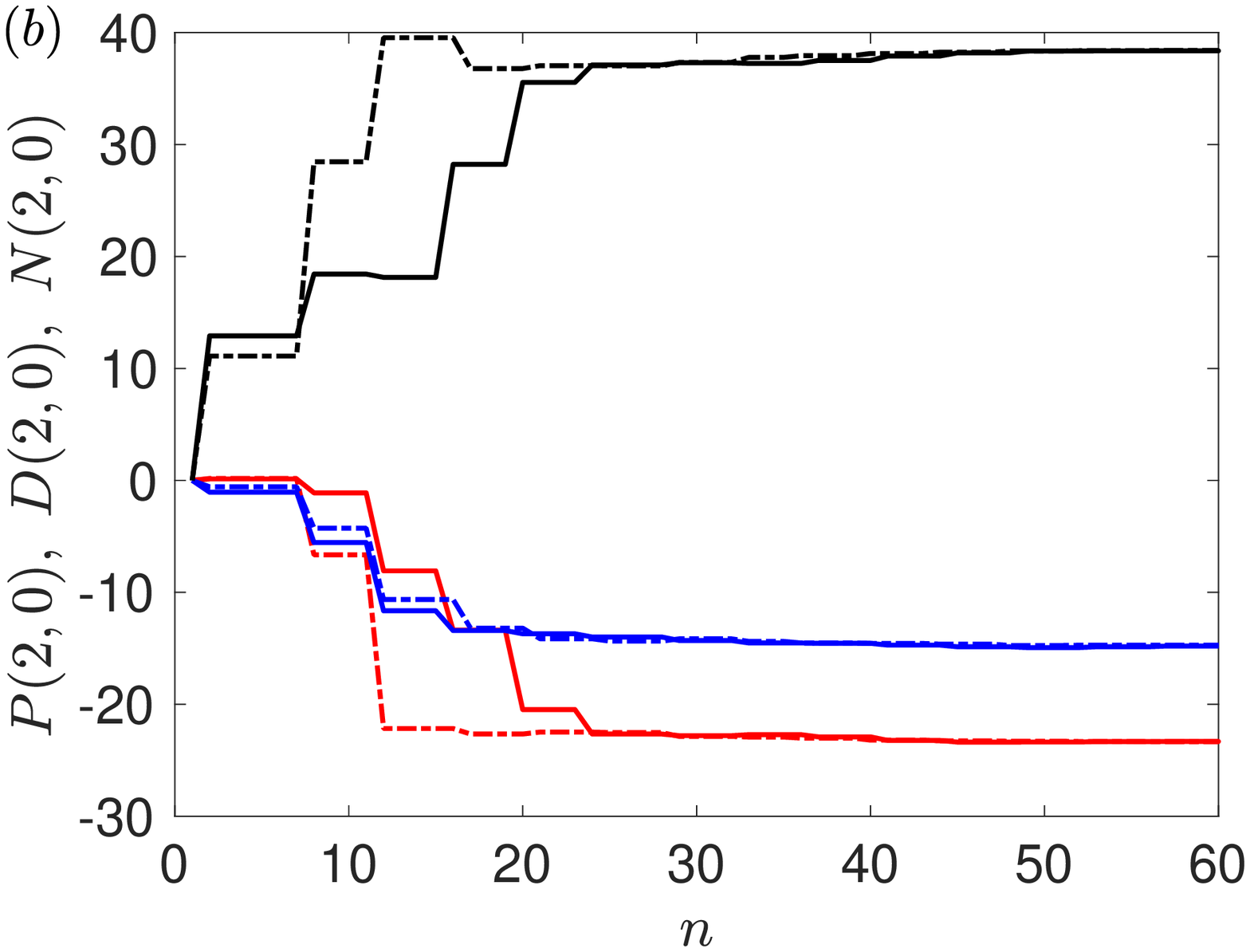}
	\includegraphics[trim = 0.1cm 0cm 1.5cm 0cm, clip,scale=0.36]{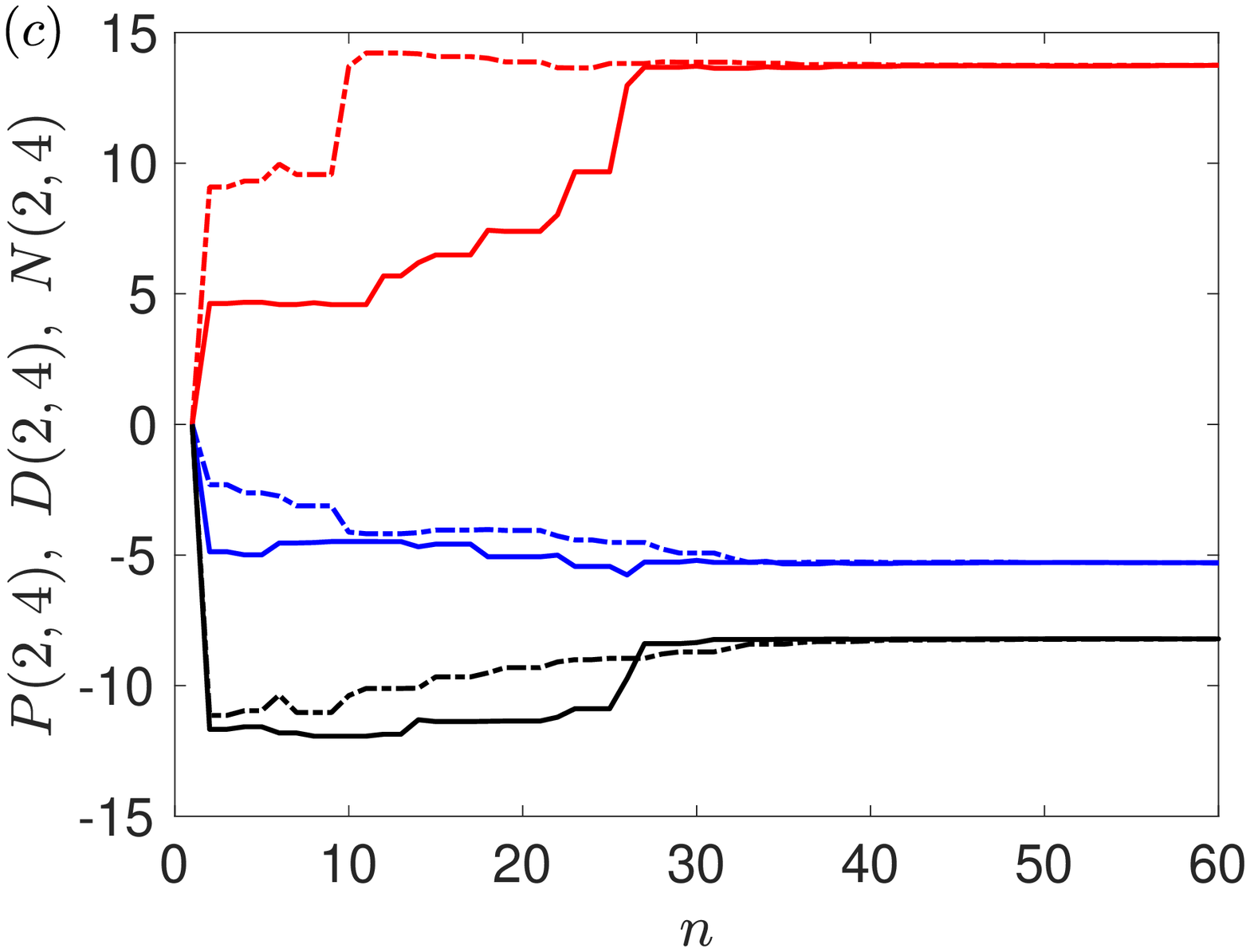}
	
	\caption{(a) Production (red), dissipation (blue) and nonlinear transfer (black) for (a) $(k_x,k_y) = (0,4)$, (b) $(k_x,k_y) = (2,0)$ and $(k_x,k_y) = (2,4)$ in P4U reconstructed from $n$ modes. Solid and dotted lines denote resolvent modes and eddy modes, respectively.}\label{fig:projection resolvent}
\end{figure}

In figure \ref{fig:projection resolvent}(a), the reconstructions are plotted for $(k_x,k_y) = (0,4)$, which is both the most energetic structure in the flow and most amplified by the resolvent. Solid and dotted lines denote resolvent and eddy modes, respectively. Production, which appears as red, requires almost 60 resolvent modes to be adequately captured. Moreover, modes 10 to 30 contribute almost zero net production and therefore modes 30 to 60 are needed. This might seem at odds with the success of the resolvent in identifying sources of production but it is actually consistent. Because the first resolvent mode is strongly biased towards production, the suboptimal modes cannot balance it with nonlinear transfer unless $\tilde{\chi}_1/\sigma_1 \ll 1$, i.e. the projection of $\hat{\boldsymbol{f}}$ onto $\hat{\boldsymbol{\phi}}_1$ is small. Therefore, the nonlinear forcing has to be heavily biased towards suboptimal modes in order to maintain an energy balance. The reconstruction using eddy modes, on the other hand, performs reasonably well with a single mode. Although 30 eddy modes are required to converge to the true value, this is considerably less than the 60 resolvent modes needed. Dissipation (blue) and nonlinear transfer (black)  converge more quickly to their true values for both sets of modes. 

Resolvent and eddy modes perform equally well as each other for $(k_x,k_y) = (2,0)$ as seen in figure \ref{fig:projection resolvent}(b). In contrast to $(0,4)$, the spanwise-constant structure has negative production and positive nonlinear transfer. The first eddy mode, which was nearly able to capture all terms in the energy budget, is no longer sufficient to reconstruct any term. Although they converge slightly faster than the resolvent modes, both sets require $n \approx 20$ in order to converge to within 5\% of their true values. 

We finally consider the oblique wave $(k_x,k_y) = (2,4)$ in figure \ref{fig:projection resolvent}(c). This structure is less straightforward than the previous two cases. The eddy modes reconstruct production with just 10 modes whereas almost 30 resolvent modes are needed. Dissipation and nonlinear transfer, on the other hand, require more than 30 eddy modes to converge to the true values while they are reconstructed with roughly the same number of resolvent modes needed to reconstruct production ($n \approx 28$). It can be concluded from figure \ref{fig:projection resolvent} that the primary benefit of adding eddy viscosity to the operator is in the reconstruction of production. For some scales, however, this might slow down reconstruction of dissipation and nonlinear transfer. 

\subsection{Role of aspect ratio} \label{sec:aspect ratio}

\begin{figure}
	\centering
	\includegraphics[trim = 0.8cm 0cm 2.5cm 0cm, clip,scale=0.3]{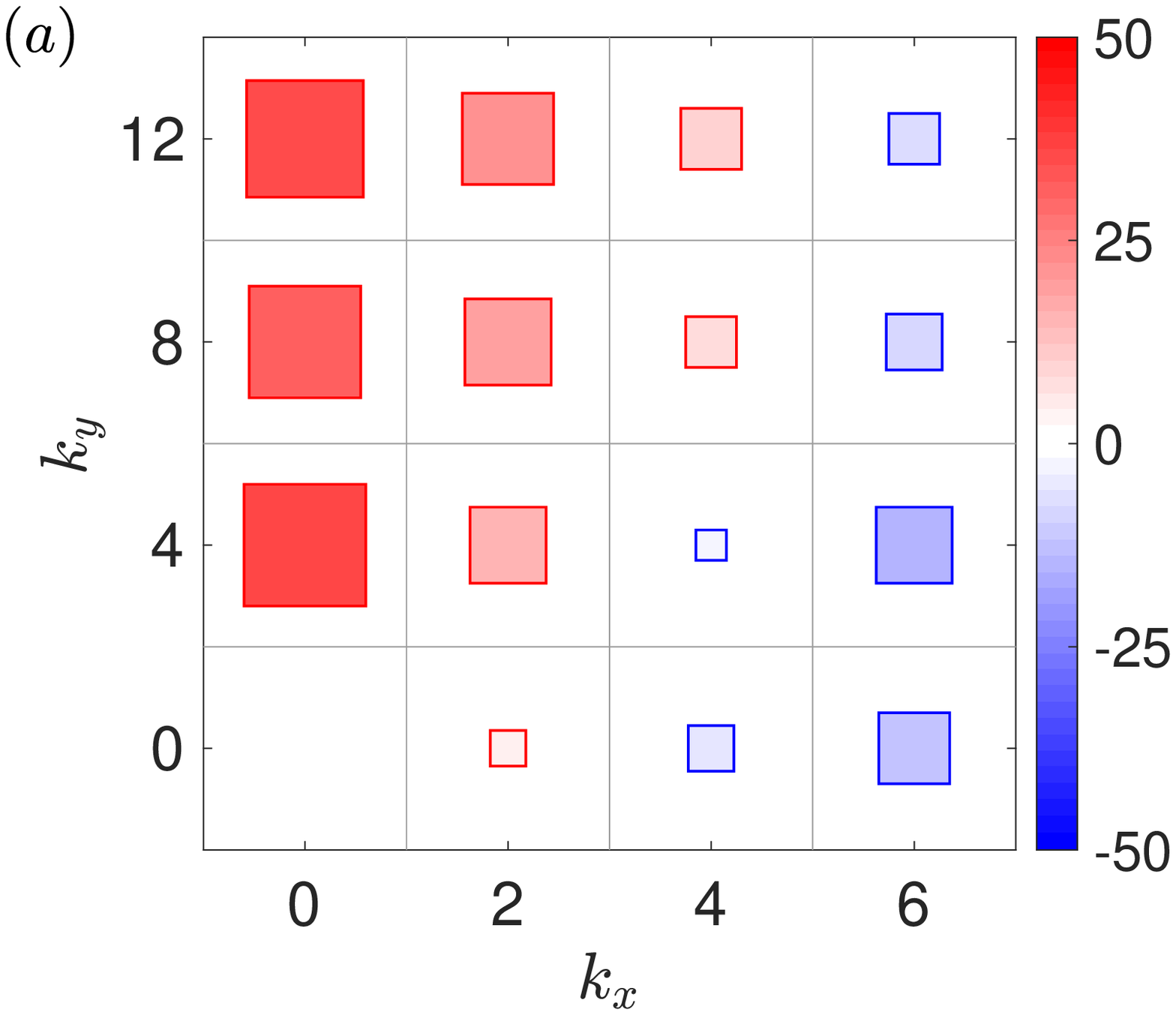}
	\includegraphics[trim = 0.8cm 0cm 2.5cm 0cm, clip,scale=0.3]{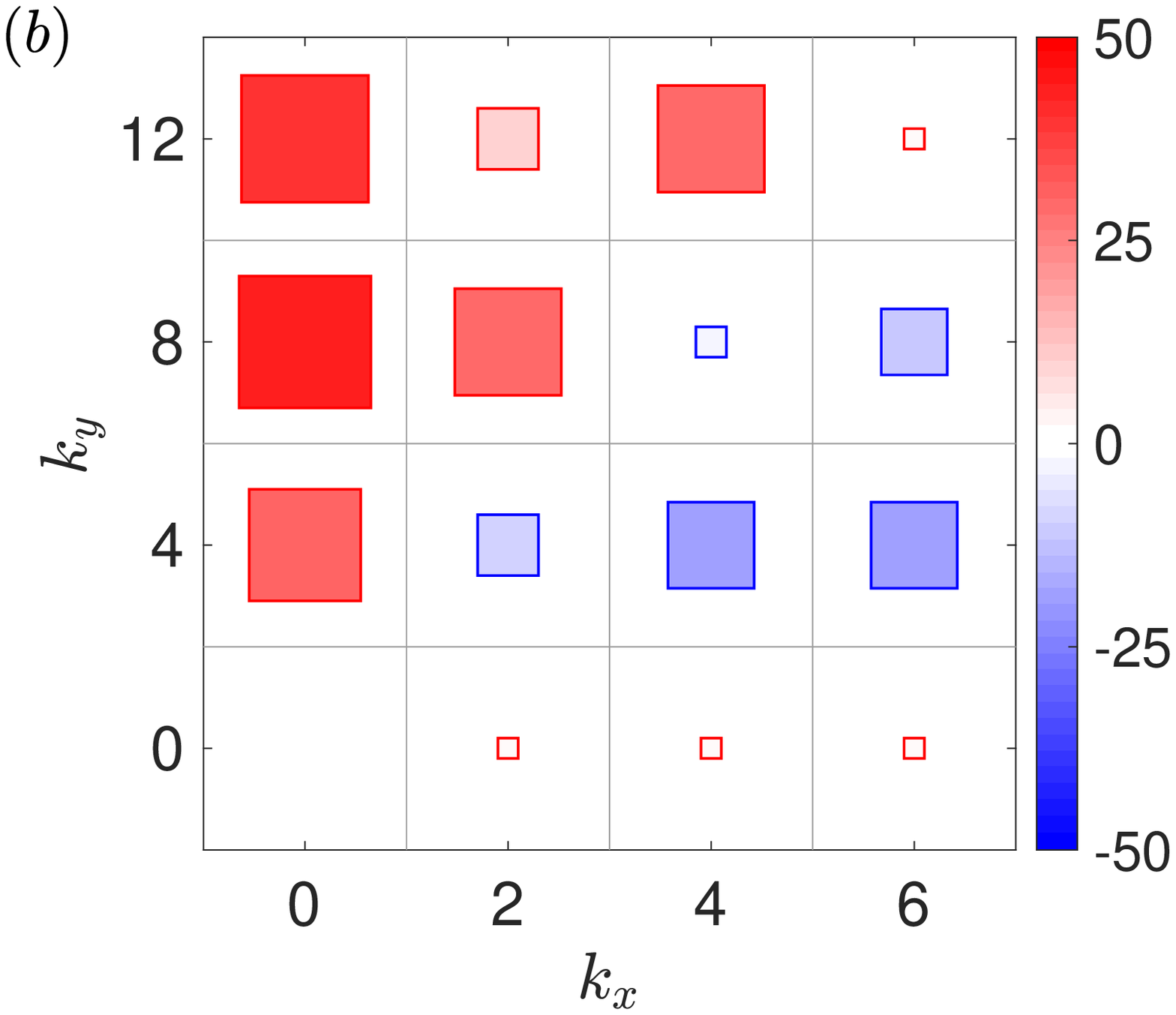}
	
	\caption{$\Delta n$ for (a) $\mathcal{T} = 25\%$ and (b) $\mathcal{T} = 1\%$. Red denotes positive values and blue negative values.}\label{fig:reconstruction error}
\end{figure}

For a more rigorous comparison between eddy and resolvent modes, we choose an error threshold $\mathcal{T}$. The number of modes $n$ is gradually increased until every term in the reconstructed energy balance is within $\mathcal{T}\%$ of the true value. We then compute the difference $\Delta n = n_{res}$ - $n_{edd}$ between the number of resolvent $n_{res}$ and eddy modes $n_{edd}$ required and present the results in figure \ref{fig:reconstruction error}. The colour red indicates that eddy modes are more efficient, or $\Delta n > 0$, and thus more resolvent modes must be included to reconstruct the energy budget. Alternatively, the colour blue designates scales for which resolvent modes are more efficient, or $\Delta n < 0$. We also assess the impact of $\mathcal{T}$ on $\Delta n$ by setting the threshold to $\mathcal{T} = 25\%$ in figure \ref{fig:reconstruction error}(a) and reducing it to $\mathcal{T} = 1\%$ in figure \ref{fig:reconstruction error}(b). 

The first observation that can be made regarding figure \ref{fig:reconstruction error} is that the choice of $\mathcal{T}$ has a relatively minor influence on the results. When $\mathcal{T}$ is reduced from 25\% to 1\%, only five tiles change sign. If we recall that most of the kinetic energy and transfer processes are among $k_x = 0$ and $k_x = 2$ scales, then the only important scale impacted by $\mathcal{T}$ is $(k_x,k_y) = (2,4)$. The reason it is positive for $\mathcal{T} = 25\%$ is that eddy modes more quickly reconstruct the dominant production term as seen in figure \ref{fig:projection resolvent}(c). It is negative for $\mathcal{T} = 1\%$ because the eddy modes take longer to reconstruct dissipation and nonlinear transfer. 

The most striking trend for both values of $\mathcal{T}$ is that the eddy modes are a more efficient basis when $k_x < k_y$. Streamwise-constant structures, in particular, are represented with more than 20 fewer modes if the velocity field is projected onto eddy modes. When we define aspect ratio as $\AR = k_y/k_x$, then $\AR = 3$ is a conservative estimate for when eddy modes are a superior basis to resolvent modes. The pattern is less clear for structures with $\AR = 2$ since the sign of $\Delta n$ is sensitive to the choice of $\mathcal{T}$. Excluding spanwise-constant structures, those with $\AR < 1$ are slightly better represented with resolvent modes although they tend to be weak energetically. 

While we only consider P4U in this section, parallel analyses for the minimal channel in appendix \ref{sec:SPOD} are consistent with the above. We opt to focus only on P4U for simplicity since the Fourier modes for a particular wavenumber pair are deterministic and travel at a single convection velocity. The details for handling statistical variability and multiple convection velocities are discussed in appendix \ref{sec:SPOD}.

\section{Conclusions} \label{sec:conclusions}

We have investigated energy transfer for the P4U ECS and low Reynolds number turbulent flow in a minimal channel. For every wavenumber pair, a balance must be achieved across production, dissipation and nonlinear transfer. For both flows, production for the energetic scales is generally positive and the largest contribution is generated by the near-wall streaks with a spanwise spacing of $\lambda_y^+ \approx 100$. Production is negative for some scales such as those that are spanwise-constant. Dissipation is negative for every scale although it is not sufficiently large to counteract production produced by the most energetic structures. As such, nonlinear transfer redistributes energy to smaller scales through the turbulent cascade and ensures each scale achieves an energy balance. The net effect of nonlinear transfer across all scales is zero but it is negative for the largest scales and positive for the rest. Spanwise-constant structures are among the largest recipients of energy through nonlinear transfer since both their production and dissipation are negative. It is interesting that they play a damping role in the minimal channel since these Tollmien-Schlichting-type waves are the first to become unstable.

Energy transfer in the DNS was compared to predictions from resolvent analysis. Similar to DNS, each mode has to satisfy a balance across production, dissipation and nonlinear transfer. Since the first resolvent mode is often representative of the true velocity field, we computed its energy balance and compared it to DNS. The first resolvent mode was successful in identifying the main production mechanisms in the flow. These are the most amplified structures by resolvent analysis and highlight the role of linear mechanisms in the sustenance of wall-bounded turbulence. For nearly all scales, production was counteracted primarily by dissipation. The nonlinear transfer, consequently, was nearly zero for every scale for resolvent analysis even though it played a major role in redistributing energy in the DNS. 

We demonstrated that nonlinear transfer could be modelled by the addition of eddy viscosity, which introduced additional dissipation into the energy balance. Its quantitative accuracy, however, was limited to the most energetic mode and we noted that it can only remove energy, suggesting it less applicable for the many scales that receive energy through nonlinear transfer. The addition of eddy viscosity, nevertheless, had an impact on the number of resolvent modes required to reconstruct the energy budget. To explore this in greater detail, each term in the energy balance was reconstructed as a function of the number of correctly resolvent weighted modes. We only considered the P4U ECS since there was a unique wave speed and deterministic structure for each wavenumber pair to simplify the analysis. We determined that the eddy basis performed better for high aspect ratio structures, particularly with respect to reconstructing production. In some cases, as many as 20 fewer eddy modes than resolvent modes were needed to properly reconstruct all terms in the energy balance. The resolvent basis was slightly better for high aspect ratio structures although these were energetically less significant. A threshold of $\AR = 3$ was a conservative estimate for when eddy modes were more efficient. 

We showed that eddy viscosity improved the basis for high aspect ratio structures by counteracting non-normality. This term resulted in a trade-off between production and nonlinear transfer in the energy balance equation for the first resolvent mode. Higher non-normality resulted in more production but less nonlinear transfer. The most amplified mechanisms, which tended to be non-normal, were therefore the most poorly represented by the first resolvent mode. The damping introduced by eddy viscosity mitigated this trade-off and resulted in a better basis for the velocity field. Its refinement for higher Reynolds number flows could improve the potential of linear models for estimation and control. 

\section{Acknowledgements} \

The authors wish to thank J. S. Park and M. D. Graham, for providing the P4U solution analysed in this article. The authors are also very grateful to M. Xie and D. Chung for providing the DNS data in the case of the minimal channel. Finally, the authors  acknowledge the financial support of the Australian Research Council. 

\appendix

\section{Linear operators} \label{sec:operators}

After elimination of the pressure, the linearized Navier-Stokes equations can be rewritten for the wall-normal velocity $\hat{w}$ and wall-normal vorticity $\hat{\eta} = ik_y\hat{u}-ik_x\hat{v}$). The matrices $\boldsymbol{A}$, $\boldsymbol{B}$, and $\boldsymbol{C}$ that appear in (\ref{eq:OSSQ}) are
\begin{subequations}
	\begin{equation}
	\boldsymbol{A} = \boldsymbol{M}\left[\begin{array}{cc} \mathcal{L}_{OS} & 0 \\ -ik_y U' & \mathcal{L}_{SQ} \end{array} \right],
	\end{equation}
	\begin{equation}
	\boldsymbol{B} = \boldsymbol{M} \left[\begin{array}{ccc} -i k_x \mathcal{D} & -i k_y \mathcal{D} & -k^2 \\ ik_y & -ik_x & 0 \end{array} \right],
	\end{equation}
	\begin{equation}
	\boldsymbol{C} = \frac{1}{k^2}\left[\begin{array}{cc} ik_x \mathcal{D} & -ik_y \\ ik_y \mathcal{D} & ik_x \\ k^2 & 0 \end{array} \right].
	\end{equation}
\end{subequations}
Both $\mathcal{D}$ and $'$ represent differentiation in the wall-normal direction and $k^2 = k_x^2+ k_y^2$. The mass matrix $\boldsymbol{M}$ is defined as
\begin{equation}
\boldsymbol{M}(k_x,k_y) = \left[\begin{array}{cc} \Delta^{-1} & 0 \\ 0 & \boldsymbol{I} \end{array} \right],
\end{equation}
where $\Delta = \mathcal{D}^2-k^2$ and $\boldsymbol{I}$ is the identity matrix. The Orr-Sommerfeld $\mathcal{L}_{OS}$ and Squire $\mathcal{L}_{SQ}$ operators are 
\begin{subequations}
	\begin{equation}
	\mathcal{L}_{OS} = -ik_xU\Delta + ik_xU'' + (1/Re_{\tau})\Delta^2,
	\end{equation}
	\begin{equation}
	\mathcal{L}_{SQ} = -ik_xU + (1/Re_{\tau}) \Delta.
	\end{equation}
\end{subequations}
With the addition of eddy viscosity, they become
\begin{subequations}
	\begin{equation}
	\mathcal{L}_{OS} = -ik_xU\Delta + ik_xU'' + \nu_T\Delta^2 + 2 \nu_T'\mathcal{D}\Delta + \nu_T''(\mathcal{D}^2 + k^2),
	\end{equation}
	\begin{equation}
	\mathcal{L}_{SQ} = -ik_xU + \nu_T \Delta + \nu_T'\mathcal{D}.
	\end{equation}
\end{subequations}

\section{Non-normality in the minimal channel case} \label{sec:SPOD}

As alluded to in \S\ref{sec:aspect ratio}, analysis of the minimal channel is complicated by the fact that each wavenumber pair has a distribution of energetic temporal frequencies. For the sake of simplicity, we choose to analyse the most energetic temporal frequency for each wavenumber pair only. The second complication is that, unlike the P4U case, there is no deterministic mode to describe the velocity field or nonlinear forcing at each $\boldsymbol{k}$ in the minimal channel. As discussed by \cite{Towne18}, this stems from the statistical variability in turbulent flows. Instead, we can replace $\hat{\boldsymbol{u}}(\boldsymbol{k})$ in (\ref{eq:projection}) with the most energetic mode from spectral proper orthogonal decomposition (SPOD) \citep{Lumley70,Picard00} which optimally represents the space-time flow statistics for any $\boldsymbol{k}$. 

The SPOD modes are computed using the procedure described in \cite{Towne18} and \cite{Muralidhar19}; a brief summary is provided here. Following Welch's method \citep{Welch67}, the DNS data for a particular $(k_x,k_y)$ are split into overlapping segments containing 128 snapshots with 50\% overlap and are Fourier-transformed in time. The data are arranged into the matrix
\begin{equation}
\hat{\boldsymbol{Q}}(\boldsymbol{k}) = \left[\begin{array}{cccc} \hat{\boldsymbol{q}}_{\omega}^{(1)} & \hat{\boldsymbol{q}}_{\omega}^{(2)} & \cdots & \hat{\boldsymbol{q}}_{\omega}^{(s)}  \end{array} \right] , 
\end{equation}
where each row represents a different temporal frequency contained in the $s$th segment. The cross-spectral density matrix for a specific wavenumber triplet $\hat{\boldsymbol{S}}(\boldsymbol{k})$ is 
\begin{equation}
\hat{\boldsymbol{S}}(\boldsymbol{k}) = \hat{\boldsymbol{Q}}(\boldsymbol{k})\hat{\boldsymbol{Q}}^*(\boldsymbol{k})  .
\end{equation} 
The SPOD eigenvectors $\hat{\boldsymbol{V}}(\boldsymbol{k})$ and eigenvalues $\boldsymbol{\Lambda}(\boldsymbol{k})$ can be obtained via an eigenvalue decomposition of the cross-spectral density matrix
\begin{equation}
\hat{\boldsymbol{S}}(\boldsymbol{k})\hat{\boldsymbol{V}}(\boldsymbol{k}) = \hat{\boldsymbol{V}}(\boldsymbol{k})\boldsymbol{\Lambda}(\boldsymbol{k}).
\end{equation}

\begin{figure}
	\centering
	\includegraphics[trim = 0.1cm 0cm 1.5cm 0cm, clip,scale=0.24]{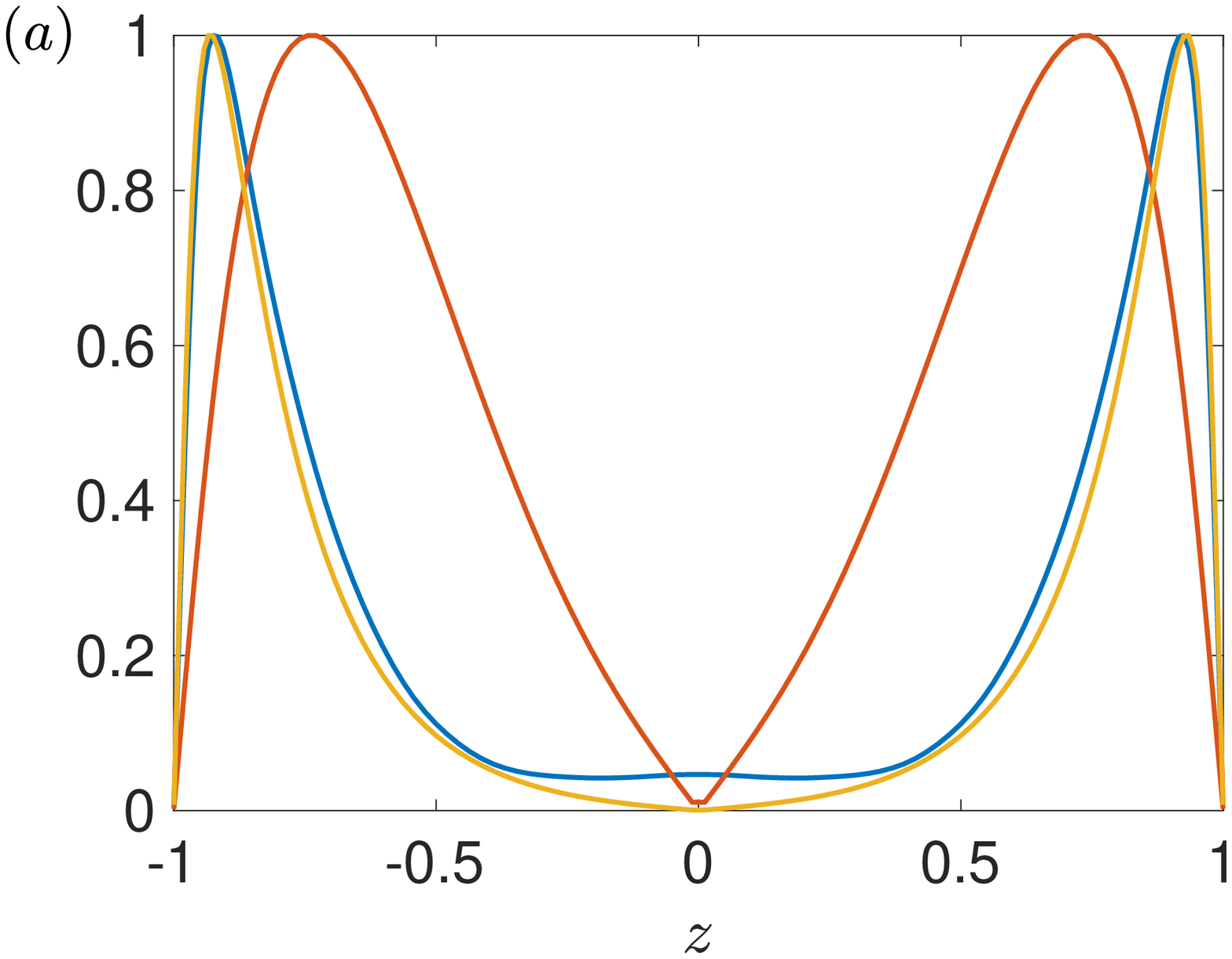}
	\includegraphics[trim = 0.1cm 0cm 1.5cm 0cm, clip,scale=0.24]{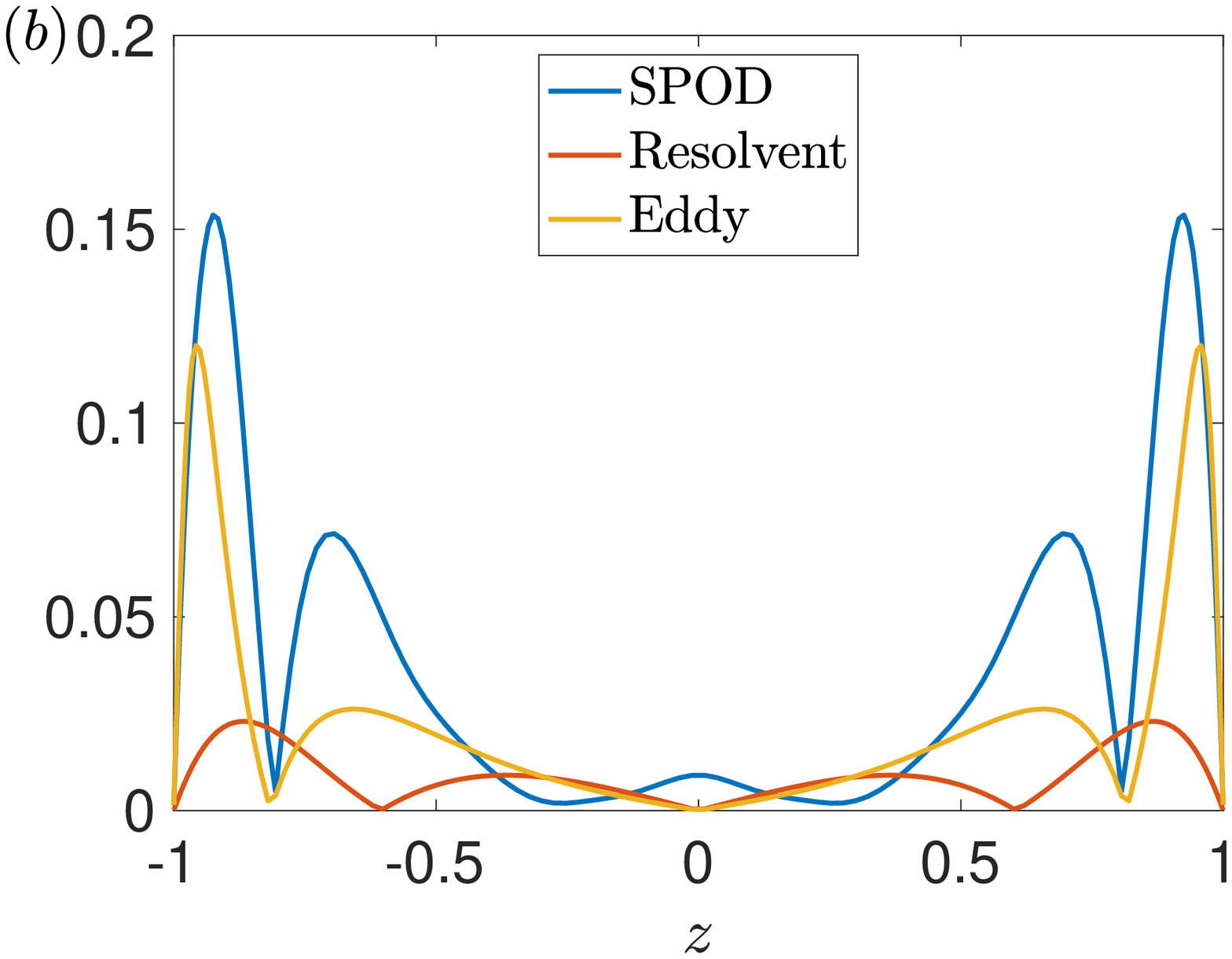}
	\includegraphics[trim = 0.1cm 0cm 1.5cm 0cm, clip,scale=0.24]{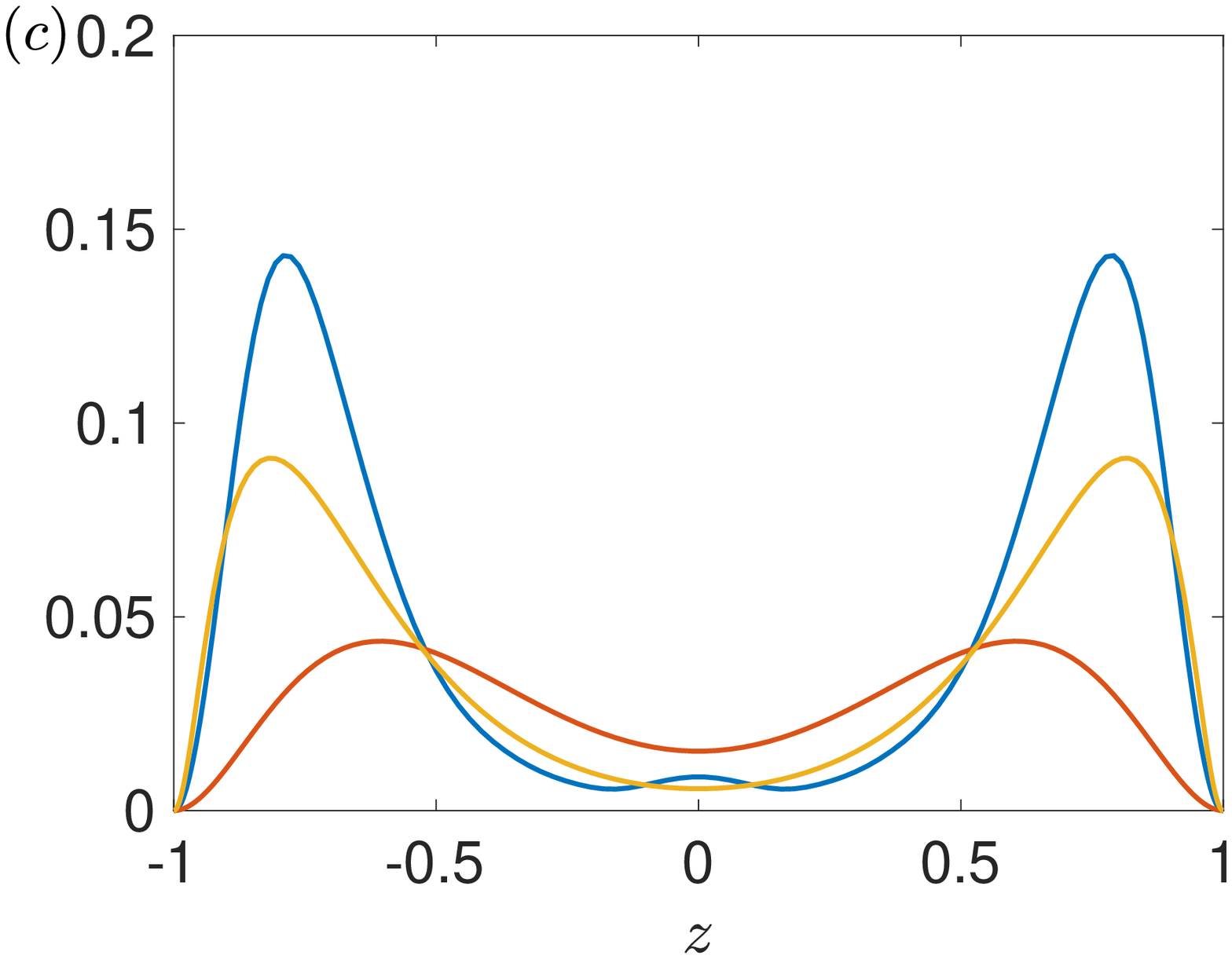}
	
	\caption{The (a) streamwise, (b) spanwise and (c) wall-normal velocity components of $\boldsymbol{k} = (0,8,0)$ in the minimal channel. The first resolvent mode is in blue, the first eddy mode is in orange, and the DNS is in yellow. The modes are normalised by the peak value of the streamwise velocity component.}\label{fig:modes minchan}
\end{figure}

In figure \ref{fig:modes minchan}, we compare the most energetic SPOD, resolvent and eddy modes for $\boldsymbol{k} = (0,8,0)$. The streamwise component of SPOD and the eddy mode are nearly identical whereas the resolvent mode predicts a much wider structure. The key improvement, however, is observed for the spanwise and wall-normal components. While the SPOD and eddy modes are not equivalent, the eddy mode is much closer than the resolvent mode in terms of magnitude. Without the presence of eddy viscosity, the first resolvent mode is more biased towards the streamwise velocity component. 

\begin{figure}
	\centering
	\includegraphics[trim = 0.1cm 0cm 1.2cm 0cm, clip,scale=0.36]{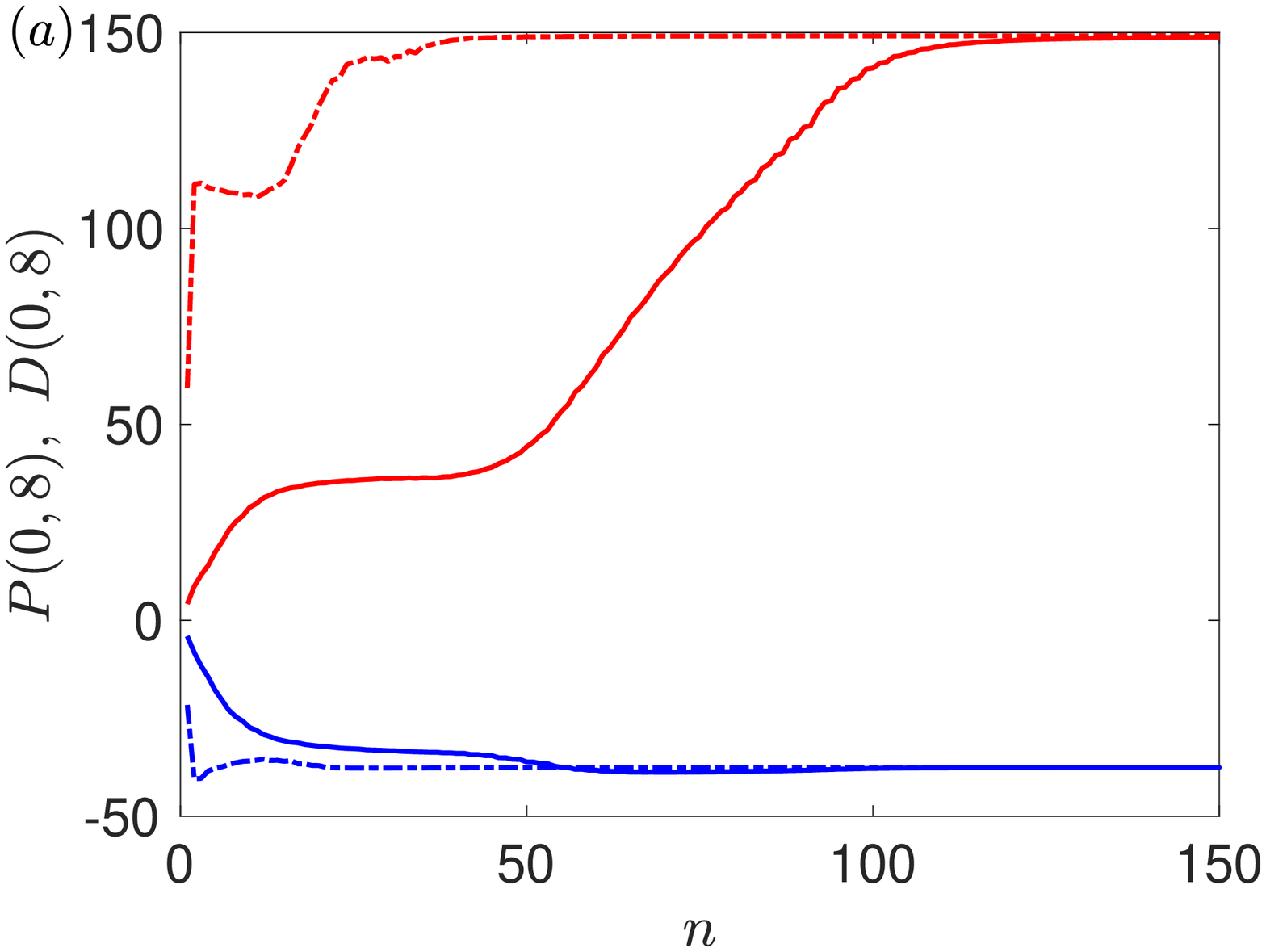}
	\includegraphics[trim = 0.1cm 0cm 1.2cm 0cm, clip,scale=0.36]{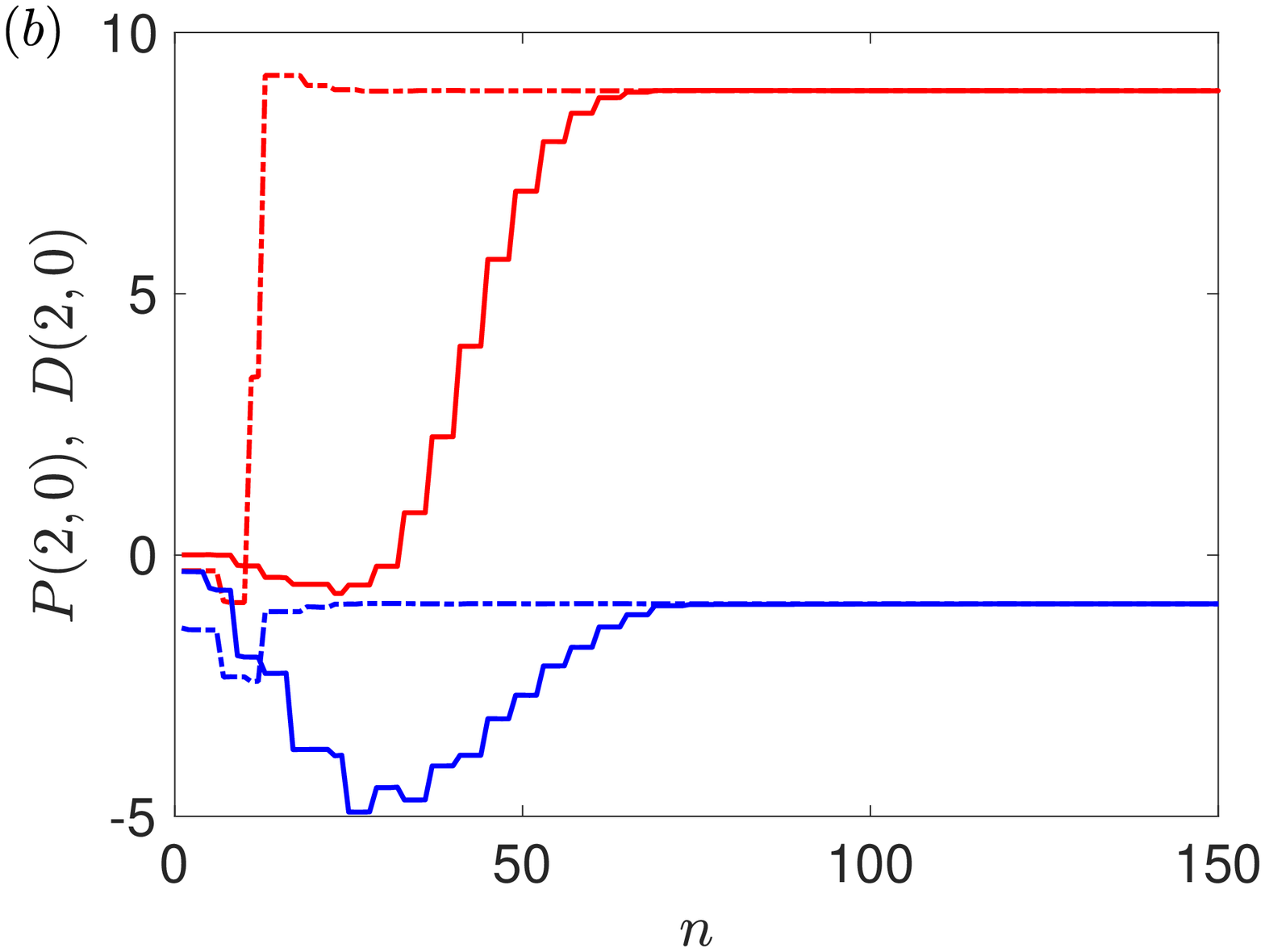}
	\includegraphics[trim = 0.1cm 0cm 1.2cm 0cm, clip,scale=0.36]{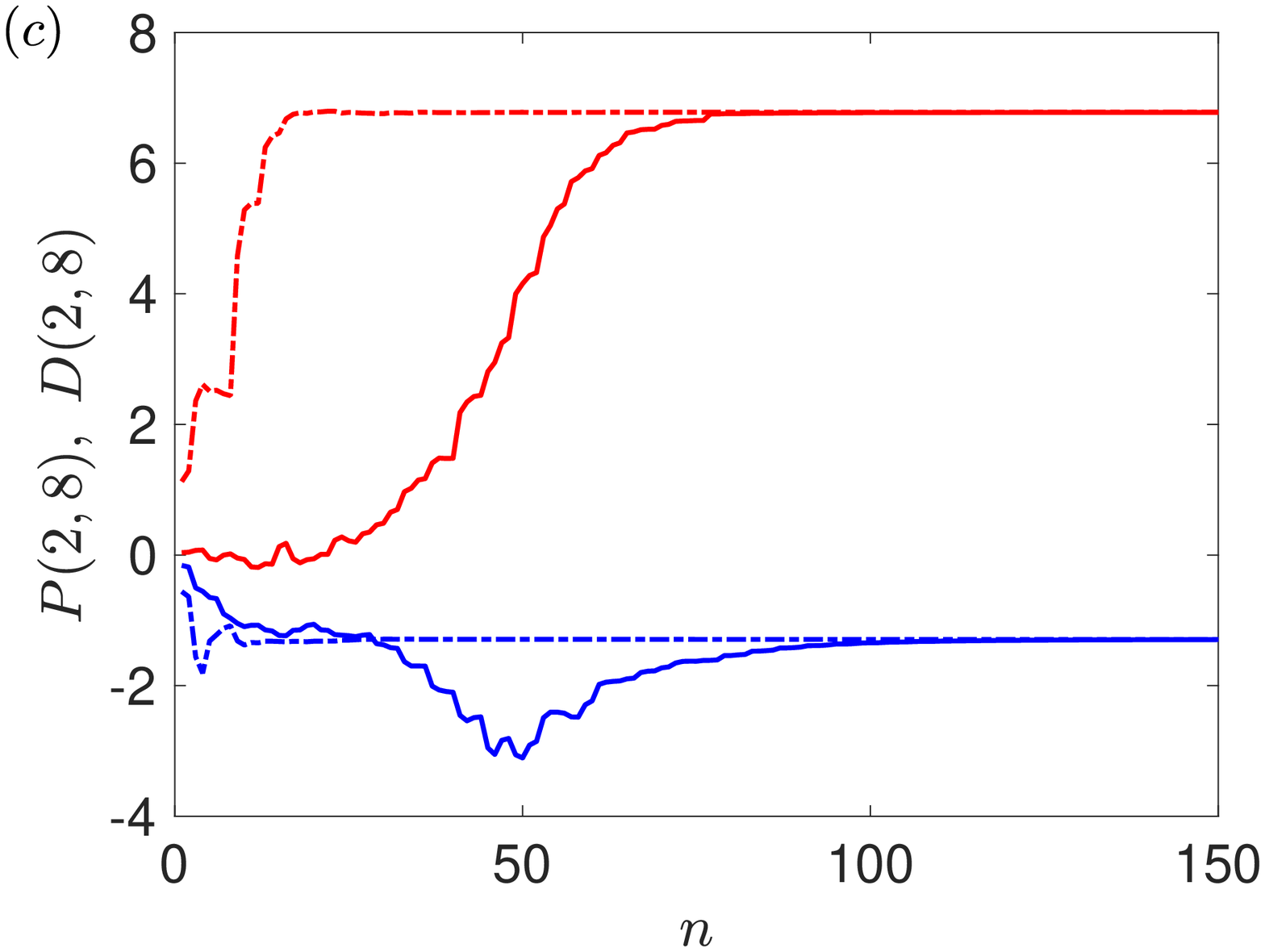}
	
	\caption{Production (red) and dissipation (blue) for (a) $(k_x,k_y) = (0,8)$, (b) $(k_x,k_y) = (2,0)$ and (c) $(k_x,k_y) = (2,8)$ in the minimal channel reconstructed from $n$ modes. Solid and dotted lines denote resolvent modes and eddy modes, respectively.}\label{fig:projection minchan}
\end{figure}

Figure \ref{fig:projection minchan} shows the resolvent reconstructions for production and dissipation only. Nonlinear transfer is not included since it would require performing SPOD for the nonlinear forcing in addition to the velocity field. Production is the term of interest since it is slowest to converge for P4U. The wavenumber pairs considered in figure \ref{fig:projection minchan} reinforce that this observation is not specific to P4U. The resolvent reconstruction of the near wall streaks in figure \ref{fig:projection minchan}(a) is particularly slow for production, requiring over 100 modes. Similar to the P4U results, a plateau region emerges around $20 < n < 50$ where the addition of resolvent modes has virtually no impact on production. Dissipation, on the other hand, converges around $n = 50$. Only 40 eddy modes, meanwhile, are needed to reconstruct the energy budget. In fact, a substantial portion is captured by the first pair of eddy modes alone, which is consistent with their close resemblance to the SPOD modes in figure \ref{fig:modes minchan}.

The reconstructions for $(k_x,k_y) = (2,0)$ and $(k_x,k_y) = (2,8)$ in figures \ref{fig:projection minchan}(b) and (c), respectively, reinforce the efficiency of eddy modes in reconstructing the energy budget. Nevertheless, the first pairs of eddy modes for these two scales are much less effective than they were for $(k_x,k_y) = (0,8)$. 

\begin{figure}
	\centering
	\includegraphics[trim = 0.8cm 0cm 2cm 0cm, clip,scale=0.3]{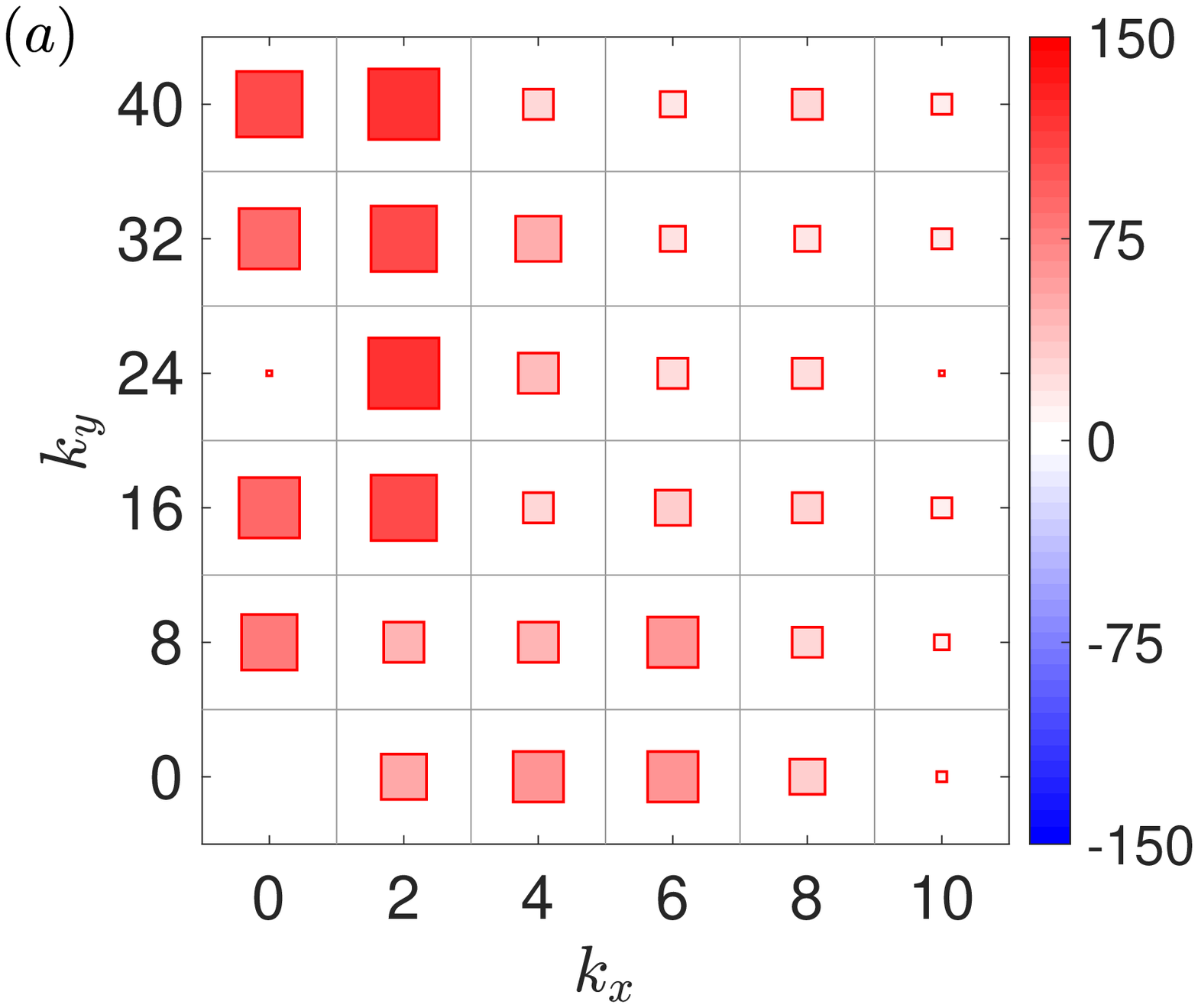}
	\includegraphics[trim = 0.8cm 0cm 2cm 0cm, clip,scale=0.3]{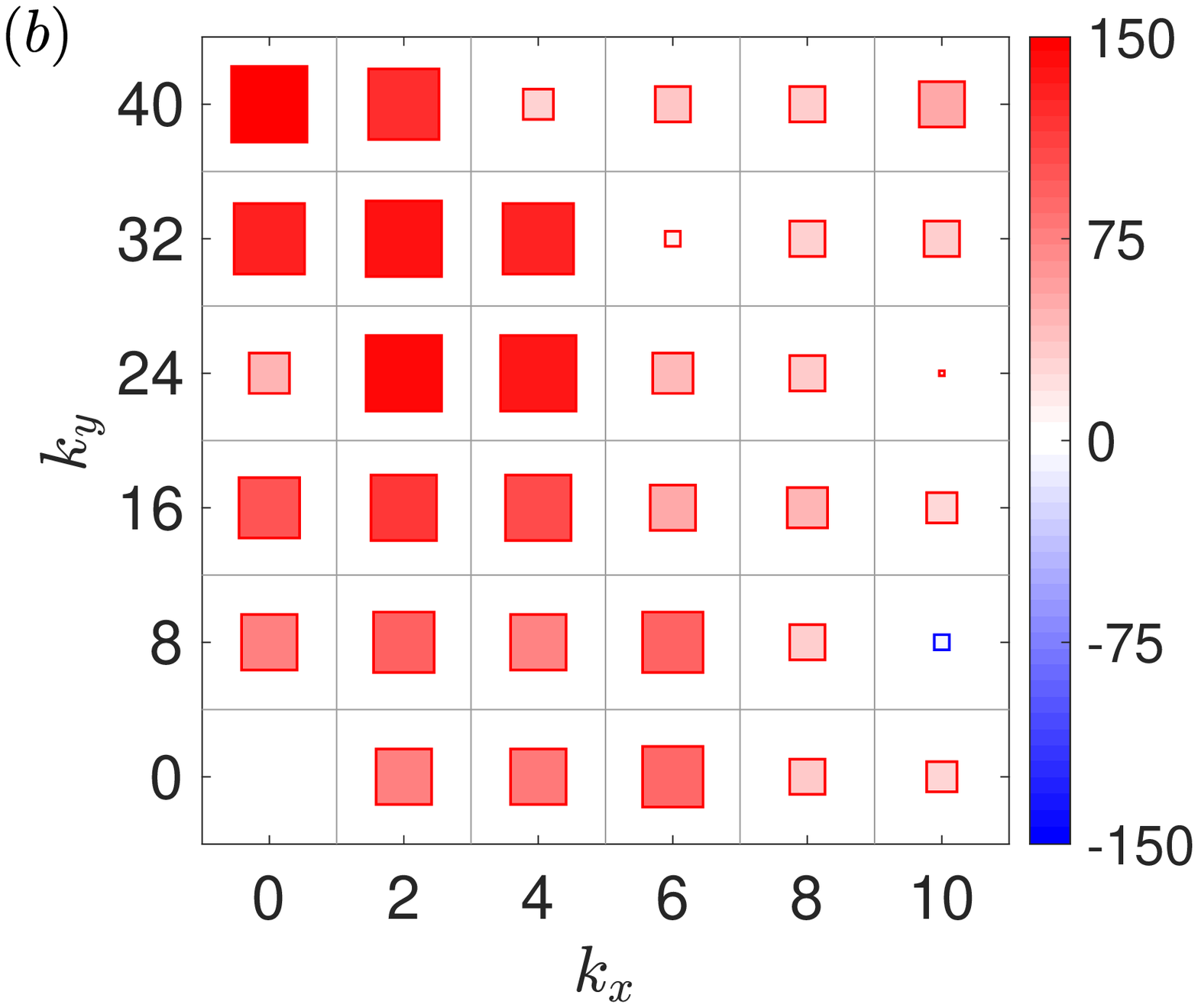}
	
	\caption{$\Delta n$ for (a) $\mathcal{T} = 25\%$ and (b) $\mathcal{T} = 1\%$. Red denotes positive values and blue negative values.}\label{fig:minchan compare}
\end{figure}

Finally, we consider $\Delta n$ for the minimal channel case in figure \ref{fig:minchan compare}. Eddy modes outperform resolvent modes for nearly every wavenumber pair regardless of the threshold $\mathcal{T}$ chosen. Due to the dimensions of the computational domain, however, $\AR > 2$ for nearly every scale that appears in figure \ref{fig:minchan compare}. Based on the P4U results, therefore, it is expected that eddy modes will be more efficient. 

\bibliographystyle{jfm}
\bibliography{energy_transfer}

\end{document}